\documentclass{JHEP3}
\usepackage{epsfig,amsfonts,amsmath,amsthm}

\newcommand{\fref}[1]{Figure~\ref{#1}}
\newcommand{\cref}[1]{Chapter~\ref{#1}}
\newcommand{\beq}{\begin{equation}}
\newcommand{\eeq}{\end{equation}}
\newcommand{\ba}{\begin{array}}
\newcommand{\ea}{\end{array}}
\newcommand{\bcenter}{\begin{center}}
\newcommand{\ecenter}{\end{center}}

\def\C{\mathbb{C}}
\def\IC{\mathbb{C}}

\def\IGa{\relax\hbox{${\rm I}\kern-.18em\Gamma$}}

\def\IR{\mathbb{R}}

\def\Z{\mathbb{Z}}
\def\IZ{\mathbb{Z}}

\def\ch{{\rm ch}}

\def\Hom{{\rm Hom}}
\def\dim{{\rm dim}}

\def\smiley{\hbox{\large$\bigcirc$\hspace{-0.80em}\raise.2ex
\hbox{$\cdot\cdot$}\kern-.61em\lower.2ex\hbox{\scriptsize$\smile$}}\ }
\def\frowny{\hbox{\large$\bigcirc$\hspace{-0.80em}\raise.2ex
\hbox{$\cdot\cdot$}\kern-.635em\lower.2ex\hbox{\scriptsize$\frown$}}\ }

\makeatletter
\let\hangafter\@hangfrom
\makeatother

\newtheorem{thm}{Theorem}[subsection]
\newcommand{\btheorem}{\begin{thm}}
\newcommand{\etheorem}{\end{thm}}
\newtheorem{lem}[thm]{Lemma}
\newcommand{\blemma}{\begin{lem}}
\newcommand{\elemma}{\end{lem}}
\newtheorem{dfn}[thm]{Definition}
\newcommand{\bdefn}{\begin{dfn}}
\newcommand{\edefn}{\end{dfn}}
\newtheorem{cor}[thm]{Corollary}
\newcommand{\bcor}{\begin{cor}}
\newcommand{\ecor}{\end{cor}}
\def\bproof{\begin{proof}} 
\def\eproof{\end{proof}}

\newcommand{\be}{\begin{equation}}
\newcommand{\ee}{\end{equation}}
\newcommand{\bea}{\begin{eqnarray}}
\newcommand{\eea}{\end{eqnarray}}
\newcommand{\bean}{\begin{eqnarray*}}
\newcommand{\eean}{\end{eqnarray*}}
\newcommand{\bc}{\begin{center}}
\newcommand{\ec}{\end{center}}
\newcommand{\comment}[1]{}

\DeclareFontFamily{U}{rsf}{} \DeclareFontShape{U}{rsf}{m}{n}{  <5> <6> rsfs5 <7> <8> <9> rsfs7 <10-> rsfs10}{}
\DeclareMathAlphabet\Scr{U}{rsf}{m}{n} \DeclareMathAlphabet\mathbi{U}{cmr}{bx}{it}

\def\C{{\mathbb C}} 
 \def\Z{{\mathbb Z}}

\def\Hom{\operatorname{Hom}} 	
\def\Ext{\operatorname{Ext}}

\def\rk{\operatorname{rk}}	\def\dim{\operatorname{dim}}

\def\ch{\operatorname{\mathrm{ch}}}

\def\ses#1#2#3{\xymatrix@1{0 \ar[r] & #1 \ar[r] & #2 \ar[r] & #3 \ar[r] & 0}}

\def\IR{\mathbb{R}}
\def\IC{\mathbb{C}}
\def\IZ{\mathbb{Z}}

\def\cale{\mathcal{E}}

\def\calo{\mathcal{O}}

\def\Ext{\mbox{Ext}}
\def\Hom{\mbox{Hom}}

\def\ch{\mbox{ch}}

\def\be{\begin{equation}}
\def\ee{\end{equation}}

\preprint{MIT-CTP-3717 \\ {\tt hep-th/0602041}}

\title{Brane Tilings and Exceptional Collections}

\author{Amihay Hanany$^1$, Christopher P. Herzog$^2$, David Vegh$^1$
\\
~\\
$^1$ Center for Theoretical Physics,
Massachusetts Institute of Technology,\\
Cambridge, MA 02139, USA.\footnote{ Research supported in part by the CTP and the LNS of MIT and
the U.S. Department of Energy under cooperative agreement $\#$DE-FC02-94ER40818. AH is also
supported in part by the BSF American--Israeli Bi--National Science Foundation and a DOE OJI
award.} \vskip 0.2cm
$^2$ Department of Phsyics, University of Washington,\\
Seattle, WA 98195, USA.\footnote{ Research supported in part by the U.S. Department
    of Energy under Grant No.~DE-FG02-96ER40956.}
\vskip 0.1cm
\email{hanany@mit.edu}, \ \email{herzog@u.washington.edu}, \ \email{dvegh@mit.edu}
}

\abstract{

Both brane tilings and exceptional collections are useful tools for describing the low energy
gauge theory on a stack of D3--branes probing a Calabi--Yau singularity.  We provide a
dictionary that translates between these two heretofore unconnected languages. Given a brane
tiling, we compute an exceptional collection of line bundles associated to the base of the
non--compact Calabi--Yau threefold.  Given an exceptional collection, we derive the periodic
quiver of the gauge theory which is the graph theoretic dual of the brane tiling. Our results
give new insight to the construction of quiver theories and their relation to geometry. }

\begin{document}


\section{Introduction}


Determining the low energy gauge theory on a stack of D--branes probing a Calabi--Yau
singularity is an important, interesting, and in general unsolved problem.  These D--brane
constructions can be used to build flux vacua in string theory, and they play an important role
in the AdS/CFT correspondence, where they yield a geometric understanding of strongly coupled
gauge theories.  While much progress has been made in understanding orbifold, toric, and other
simple Calabi--Yau singularities, the general case remains elusive.

Two of the most powerful techniques for unearthing these gauge theories are the brane tiling
method pioneered by \cite{Hanany:2005ve, Franco:2005rj, Franco:2005sm} and exceptional
collections first mentioned in the AdS/CFT context in \cite{Cachazo:2001sg}.  The relation
between these two methods has up to this point remained obscure.  In this paper, we show how to
translate one language into the other.

More specifically, we have in mind D3--branes in type IIB string theory.  The ten dimensional
geometry is divided up into a Minkowski part ${\mathbb R}^{3,1}$ which the D3--branes occupy and
a transverse Calabi--Yau threefold $Y$.  Placing the D3--branes at a singularity of $Y$ produces
complicated quiver gauge theories which preserve ${\mathcal N}=1$ supersymmetry.

One of the best features of the brane tiling method is the ease with which the superpotential of
the quiver gauge theory can be extracted.  A brane tiling is a bipartite tiling of the torus
$T^2$, and the superpotential terms are just the nodes of this tiling with coefficient $\pm 1$
given by the coloring of the node.  No other method of relating gauge theory to geometric
singularity has as yet produced such a simple way of extracting the superpotential.

For the brane tiling method to work, one starts with a toric Calabi--Yau three--fold
singularity.  The toric condition means that $Y$ possesses three $U(1)$ isometries. There are
countably many interesting toric Calabi--Yau singularities, but the toric condition is a
substantial restriction on $Y$. By using brane tilings, older algorithms (\cite{Feng:2000mi,
Feng:2001xr}) get vastly simplified and reinterpreted.

For the exceptional collection method to work, one needs to be able to resolve partially the
Calabi--Yau singularity by blowing up a complex surface -- the exceptional collection lives on
this surface.  There are many both toric and non--toric Calabi--Yau singularities which can be
resolved in this manner. The exceptional collection method was in large part developed to study
some simple non--toric singularities, the non--toric del Pezzos \cite{Wijnholt:2002qz}.

While the superpotential can be extracted from an exceptional collection, the process is more
abstract and less intuitive than for the brane tiling.  In the exceptional collection case,
deriving the superpotential requires working with $A$--infinity algebras \cite{Aspinwall:2004bs,
Aspinwall:2005ur}.

The exceptional collection method as applied to deriving quiver gauge theories rests on
relatively firm mathematical and physical foundations \cite{Herzog:2004qw, Aspinwall:2004vm,
Herzog:2005sy, Bergman:2005kv, Herzog:2003dj}. From the perspective of the topological B--model,
the objects in the collection can be understood as a nice basis of D--branes and the maps
between the objects as massless open strings.

The brane tiling method began as an extremely remarkable observation: the tiling contains all
the information of the quiver gauge theory, and hence proves to be a very useful tool in its
study and construction. The toric diagram  of the Calabi--Yau manifold can be easily obtained by
either computing the determinant of the Kasteleyn matrix or by determining the zig--zag paths.
Recent results  \cite{Hanany:2005ss} allow for computing the tiling directly from the toric
diagram.
More recently, the paper of \cite{Feng:2005gw} gave a physical interpretation of the dimer model
as a tiling of D6--branes in the mirror topological A--model.

By providing a translation between the brane tiling and the exceptional collection, we put the
brane tiling, along with its easy superpotential calculation, on a firmer mathematical and
physical footing.  Our results fall short of a general proof that the brane tiling method is
equivalent to exceptional collections for toric Calabi--Yau singularities.  Instead, we provide
a well motivated conjecture of the way this map will work which we can prove example by example.
By relating the tiling to exceptional collections which are topological B--model objects, our
approach is complementary to that of \cite{Feng:2005gw}.

In order for our translation between the brane tiling and the exceptional collection to work, we
henceforth restrict to toric Calabi--Yau threefold singularities which can be partially resolved
by blowing up a complex surface.

In the next section, we begin by reviewing some elementary material about quivers, quiver gauge
theories, and toric geometry.  Section 3 contains a review of the brane tiling method.  The
principal results of the paper are contained in Sections 4 and 5.

Section 4 contains a brief review of the exceptional collection method and a map from the
exceptional collection to the brane tiling.  We argue that the periodic quiver which is the dual
graph of the brane tiling can be constructed from a consideration of Wilson lines.

In Section 5, we proceed in the other direction, mapping the brane tiling onto an exceptional
collection. The cornerstone of this mapping is the realization that internal perfect matchings
are in one--to--one correspondence with exceptional collections of line bundles.

\section{Quivers and toric diagrams}
\label{section_quivers}

\begin{figure}[ht]
  \epsfxsize = 5cm
  \centerline{\epsfbox{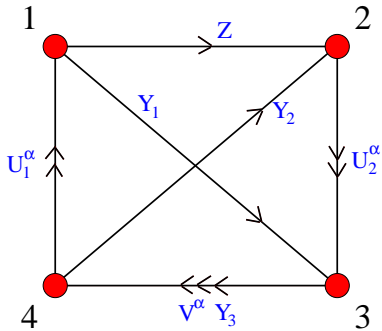}}
  \caption{Quiver of ${\bf dP}_1$.  The theory contains four $U(N)$ gauge groups labeled by the nodes
  of the quiver. The arrows label bifundamental fields transforming in the (anti--)fundamental
  representation of the groups at the endpoints.}
  \label{quiver_dp1}
\end{figure}

The matter content of the quiver gauge theory is neatly summarized in the {\bf quiver graph}
\cite{Douglas:1996sw} which also generalizes the Dynkin diagrams. Each node in the quiver (see
e.g.~\fref{quiver_dp1}) may carry an index, $N_i$, for the $i^\textrm{th}$ node and denotes a
$U(N_i)$ gauge group. The edges (arrows) label the chiral bifundamental multiplets. These fields
transform in the fundamental representation of $U(N_i)$ and in the anti--fundamental of $U(N_j)$
where $i$ and $j$ represent the nodes in the quiver that are the head and tail of the
corresponding arrow.

To be gauge anomaly free, for each gauge group, the number of chiral fermions in the fundamental
representation must equal the number in the antifundamental representation. This anomaly
cancellation means that for a fixed node in the quiver, the number of incoming and outgoing
arrows are the same.

\begin{figure}[ht]
  \epsfxsize = 6cm
  \centerline{\epsfbox{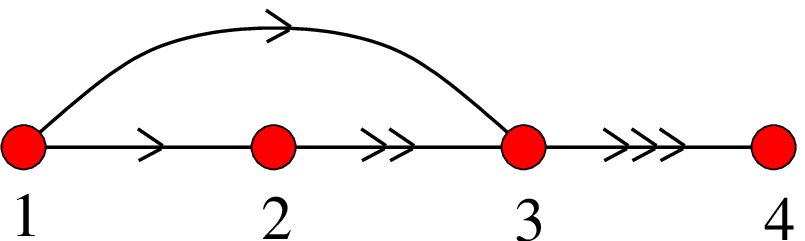}}
  \caption{${\bf dP}_1$ Beilinson quiver.}
  \label{dPbeil}
\end{figure}

By deleting certain arrows in the quiver, one obtains another graph, the so--called {\bf
Beilinson quiver}. In this quiver there exists an ordering of the nodes such that there are no
arrows pointing backwards (for an example see \fref{dPbeil}). Generically, there are many
Beilinson quivers corresponding to a given quiver. These quivers can be thought of as subquivers
that contain no oriented loops.  A more precise definition can be found in Section
\ref{section_excol}.

We are taking a small liberty with the term Beilinson quiver. Historically, Beilinson quiver
referred only to projective space (see for example \cite{Douglas:2000qw}).
  In the context of D--branes and Calabi--Yau manifolds, that would mean placing a stack of D--branes at a singularity where a ${\mathbb P}^2$
had shrunk to zero size.  The Beilinson quiver associated to ${\mathbb P}^2$ is then obtained by
eliminating backward pointing arrows in the full gauge theory quiver.  (These Beilinson quivers
are sometimes called Bondal quivers \cite{Bergman:2005mz}.)


In order to fully specify the Lagrangian, we need to give the {\bf superpotential} as well,
which is a polynomial in gauge invariant operators.  For example, for ${\bf dP}_1$ the
superpotential is
\be
\label{spot_dp1}
  W=\epsilon_{\alpha \beta} U_1^{\alpha} V^{\beta}Y_1 -
       \epsilon_{\alpha \beta} U_2^{\alpha} Y_2 V^{\beta} -
       \epsilon_{\alpha \beta} U_1^{\alpha} Y_3 U_2^{\beta} Z \ .
\ee

\begin{figure}[ht]
  \epsfxsize = 5cm
  \centerline{\epsfbox{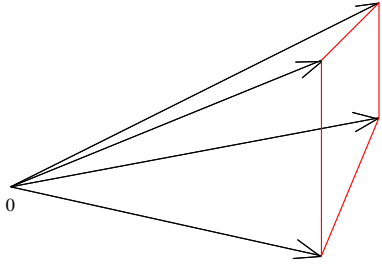}}
  \caption{The cone for the variety. The coordinates of the spanning vectors are integers.
  The endpoints are coplanar following from the Calabi--Yau condition.}
  \label{conefig}
\end{figure}

The AdS/CFT dual theory is determined by the Calabi--Yau threefold $Y$. For the purpose of this
paper, we don't need explicit metrics.  Instead, we will use toric geometry (\cite{Leung:1997tw,
fulton}) to treat the topology of these singular manifolds. To use toric methods, we restrict
the class of possible spaces to toric ones, i.e.~the isometry group of $Y$ contains a
$3$--torus. The variety can be defined by a strongly convex rational polyhedral cone $\sigma$
spanned by a set of vectors  $(\{ v_r \})$ on the integer lattice $N$ (\fref{conefig}).

The lattice is three dimensional so that we obtain a (complex) 3d space. Let
$M=\mbox{Hom}(N,\IZ)$ be the dual lattice with pairing denoted by $\langle \cdot,\cdot \rangle$.
The dual cone $\sigma^v$ is the set of vectors that are nonnegative on $\sigma$. The lattice
points in $\sigma^v$ determine a finitely generated commutative semigroup:
\be
S_{\sigma} = \sigma^v \cap M = \{ u\in M :  \langle u,v \rangle \geq 0 \ \ \mbox{for all} \ v
\in \sigma \} \ .
\ee
The corresponding commutative $\IC [ S_{\sigma} ]$ algebra defines the $U_{\sigma}$ variety by
its spectrum
\be
  U_{\sigma} = \mbox{Spec}(\IC [ S_{\sigma} ] ) \ .
\ee The so--called moment map for the torus action gives $Y$ as a
Lagrangian $T^3$ fibration over the dual cone. For details of this map the reader should refer
to \cite{fulton}.

\begin{figure}[ht]
  \epsfxsize = 2.5cm
  \centerline{\epsfbox{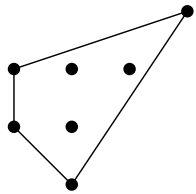}}
  \caption{The toric diagram for $L^{1,7,3}$ which is part of the recently discovered
  series of $L^{abc}$ metrics (\cite{Cvetic:2005ft, Cvetic:2005vk}). The dual quiver theories
  have been constructed in \cite{Franco:2005sm, Benvenuti:2005ja, Butti:2005sw}.}
  \label{tdL173}
\end{figure}

For each spanning vector $v_r$ there is a corresponding $D_r$ (Weil) divisor in the toric
variety. Principal divisors are of the form
\be
  \sum_r \langle m, v_r \rangle D_r \ ,
\ee
for $m \in M$. The Calabi--Yau condition states that $c_1 (Y)=0$, i.e.~the canonical class is
trivial
\be
  K = -\sum_r D_r = -\sum_i \langle m, v_r \rangle D_r \ .
\ee
The last equality implies that the endpoints of the $(\{ v_r \})$ vectors are coplanar, so with
an appropriate $SL(3,\Z)$ transformation a convex integer polygon in two dimensions can be
obtained (see e.g.~\fref{tdL173}). We will refer to this polygon as the {\bf toric diagram} of
the singularity \cite{Martelli:2004wu, Martelli:2005tp, Franco:2005sm}. Weil divisors can be
specified as integer functions over the external lattice points of the toric diagram. Principal
divisors are simply linear functions; the canonical class is a constant function.

\newpage
\section{Brane tilings}
\label{section_branetilings}


\subsection{Tilings}
\label{subsec_tilings}

In this section we give a short introduction to brane tilings \cite{Franco:2005rj}. Brane
tilings (a.k.a.~dimer graphs) have arisen in two string theory contexts:

\begin{quote}
\begin{description}
  \item[Quiver gauge theories] that are obtained by placing D3--branes at the tip of a
  non--compact toric Calabi--Yau cone
    \cite{Hanany:2005ve, Franco:2005rj, Hanany:2005ss, Feng:2005gw}, and
  \item[Topological string theory,] more specifically, the partition function
  of the topological A--model defined on the same Calabi--Yau cone \cite{Okounkov:2003sp, Iqbal:2003ds}.
\end{description}
\end{quote}

In the following, we describe tilings from the first point of view. The brane tiling is a
generalization of brane boxes \cite{Hanany:1997tb, Hanany:1998it} and brane diamonds
\cite{Aganagic:1999fe}. The tiling graph encodes the quiver and tree--level superpotential
information, thus fully specifying the 4d $\mathcal{N}=1$ quiver theory Lagrangian. The toric
diagram of the corresponding Calabi--Yau manifold can be easily computed by means of the Fast
Forward Algorithm \cite{Franco:2005rj}. On the other hand, given the toric diagram, the tiling
is simply obtained by the Fast Inverse Algorithm \cite{Hanany:2005ss, Feng:2005gw}. The
equivalence of the Forward and the Fast Forward Algorithms has recently been established in
\cite{Franco:2006gc}.

\begin{figure}[ht]
\begin{center}
  \epsfxsize = 4cm
  \centerline{\epsfbox{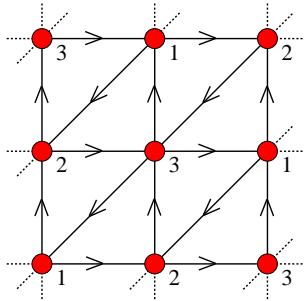}}
  \caption{The ${\mathbb P}^2$ periodic quiver. The nodes denote $U(N)$ gauge groups;
the directed edges between them are bifundamental fields. The plaquettes of the quiver graph are
terms in the superpotential. This example has three gauge groups, labeled by numbers.
Identifying nodes with the same labels (i.e.~``compactifying'' the periodic quiver) yields the
usual quiver diagram.}
  \label{infq}
\end{center}
\end{figure}

\begin{figure}[ht]
\begin{center}
  \epsfxsize = 11cm
  \centerline{\epsfbox{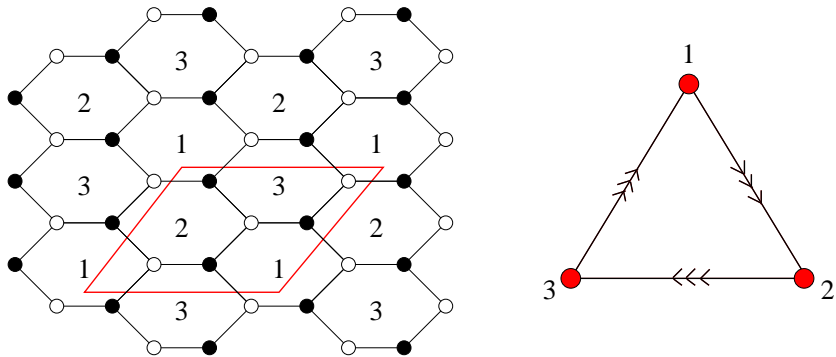}}
  \caption{${\mathbb P}^2$ brane tiling and quiver. The unit cell of the lattice is shown in red.
The theory has three gauge groups (faces in the tiling) and six cubic terms in the
superpotential (valence three nodes of the tiling).}
  \label{dP0fig1}
\end{center}
\end{figure}

The {\bf brane tiling} is a periodic bipartite graph\footnote{A planar graph is bipartite if the
nodes can be colored in black and white, such that edges only connect black nodes to white nodes
and vice versa.}
on the plane. Equivalently, one may draw it on the surface of a 2--torus. The faces label gauge
groups, the edges are chiral bifundamental fields, and the nodes are terms in the
superpotential. The dual graph of the brane tiling is the {\bf periodic quiver}. Roughly
speaking, the periodic quiver is the quiver drawn on a 2--torus such that the plaquettes give
the terms in the superpotential. \fref{infq} shows an example of a periodic quiver for the well
known case of $\C^3/\Z_3$ (otherwise known as the complex cone over ${\mathbb P}^2$). Nodes
carry three different labels and nodes with the same label are identified. The corresponding
tiling is shown in \fref{dP0fig1}.

\be
\begin{array}{cc|cc|ccc}
 \mbox{{\bf Brane tiling}} & & \
 \mbox{{\bf Periodic quiver}} & &\ \ & \mbox{{\bf Gauge theory}} \\
 \hline \hline
 \mbox{faces} & & \mbox{nodes} & & & U(N) \ \mbox{gauge groups} \\
 \mbox{edges} & & \mbox{edges} & & & \mbox{bifundamental fields} \\
 \mbox{nodes} & & \mbox{plaquettes} & & & \mbox{superpotential terms}
\end{array}
\nonumber \label{dictfig}
\ee

The tiling provides us with a simple geometrical unification of quiver and superpotential data.
The bipartite property of the tiling implies that each face in the brane tiling has an even
number of edges and that the dual quiver has an equal number of incoming and outgoing arrows for
each gauge group. As discussed in Section \ref{section_quivers}, equal numbers of incoming and
outgoing arrows are required by gauge anomaly cancellation.
%
To each term in the superpotential there is a plaquette in the periodic quiver and a black or
white node in the tiling.
The color of the node in the tiling tells us the sign of the term. Since a bifundamental field
joins a white and black node in the tiling, we conclude that each bifundamental field appears
exactly twice in the superpotential, once with a plus and once with a minus sign.

As a simple example, \fref{dP0fig1} shows the brane tiling and the quiver for ${\mathbb P}^2$.
We see that the brane tiling contains three faces; these correspond to the three gauge groups
(nodes) in the quiver. The nine edges in the tiling are the bifundamental fields. The six nodes
of the tiling immediately give the following superpotential:
\bea
\nonumber
  W=X_{12}^{(1)} X_{23}^{(2)} X_{31}^{(3)} + X_{12}^{(2)} X_{23}^{(3)} X_{31}^{(1)}+X_{12}^{(3)} X_{23}^{(1)} X_{31}^{(2)} \\
-X_{12}^{(3)} X_{23}^{(2)} X_{31}^{(1)} - X_{12}^{(2)} X_{23}^{(1)} X_{31}^{(3)}  - X_{12}^{(1)}
X_{23}^{(3)} X_{31}^{(2)} \ .
\eea
Here $X_{ij}^{(k)}$ denotes the bifundamentals going from gauge group $i$ to $j$, and $k$ labels
the different fields.

\begin{figure}[ht]
  \epsfxsize = 7cm
  \centerline{\epsfbox{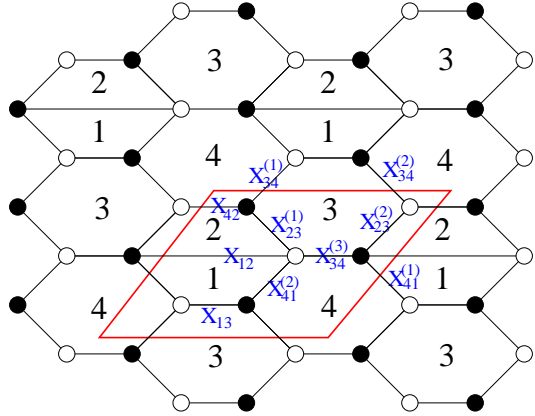}}
  \caption{Brane tiling for ${\bf dP}_1$. The fundamental cell of the periodic graph is shown
in red, the fields corresponding to the edges are shown in blue. The numbers label the faces
that correspond to the groups (nodes) in the dual quiver graph.}
\label{dp1_tiling}
\end{figure}

Another example is the {\it del Pezzo 1} ($\bf{dP}_1$) theory (\fref{dp1_tiling}). The tiling
contains four faces which label the four gauge groups.

We have already seen the corresponding quiver (\fref{quiver_dp1}). The tiling gives the
superpotential:
\bea
\nonumber
  W= X_{23}^{(1)} X_{34}^{(1)} X_{42} + X_{12} X_{23}^{(2)} X_{34}^{(3)} X_{41}^{(1)} + X_{13} X_{34}^{(2)} X_{41}^{(2)} \\
  - X_{13} X_{34}^{(1)} X_{41}^{(1)}
  - X_{12} X_{23}^{(1)} X_{34}^{(3)} X_{41}^{(2)} - X_{23}^{(2)} X_{34}^{(2)} X_{42}
\eea
which, after making the relabeling $\{ X_{42}=Y_2, \ X_{12}=Z, \ X_{13}=Y_1,\ X_{34}^{(\alpha)}
= V^{(\alpha)},\ X_{34}^{(3)}=Y_3,\ X_{23}^{(\alpha)}=U_2^{(2-\alpha)},\ \
X_{41}^{(\alpha)}=U_1^{(2-\alpha)} \}$, is the same as (\ref{spot_dp1}).

One can compute the toric diagram related to the moduli space of this theory by means of the
{\bf Kasteleyn matrix} (\ref{kastdP1}). The Kasteleyn matrix is
 the adjacency matrix of the tiling graph.  More precisely,
the rows are labeled by the black nodes, the columns by the white nodes. The corresponding entry
is zero if the two nodes are not connected; otherwise it is the appropriate weight of the
connecting edge. For details of building the Kasteleyn matrix the reader should refer to
\cite{Franco:2005rj, Hanany:2005ve, Kenyon:2002a, Kenyon:2003uj}.
\beq
  \label{kastdP1}
  K=\left( \begin{array}{ccc}
  z^{-1} & 1 & w^{-1} \\
  1 & 1-z & z \\
  w & 1 & 1
  \end{array}
  \right)
\eeq

\begin{figure}[ht]
  \epsfxsize = 2.5cm
  \centerline{\epsfbox{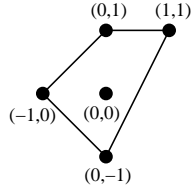}}
  \caption{Toric diagram for ${\bf dP}_1$. The multiplicity of the internal
point is four.}
  \label{dP1_td}
\end{figure}

The determinant of the Kasteleyn matrix gives the spectral curve
\be
  P(w,z) \equiv \mbox{det} K = -4 + w^{-1} + z^{-1} + z + wz \ .
\ee
The $P(w,z)=xy$ equation and its deformations describe the mirror Calabi--Yau as a fibration.
The Newton polygon of this polynomial gives the toric diagram of the threefold
(\fref{dP1_td}).\footnote{ The toric diagram can also be computed by means of the zig--zag paths
of the tiling; for details, see \cite{Hanany:2005ss}.}

\subsection{Perfect matchings}
\label{subsec_pms}

\begin{figure}[ht]
  \epsfxsize = 14.2cm
  \centerline{\epsfbox{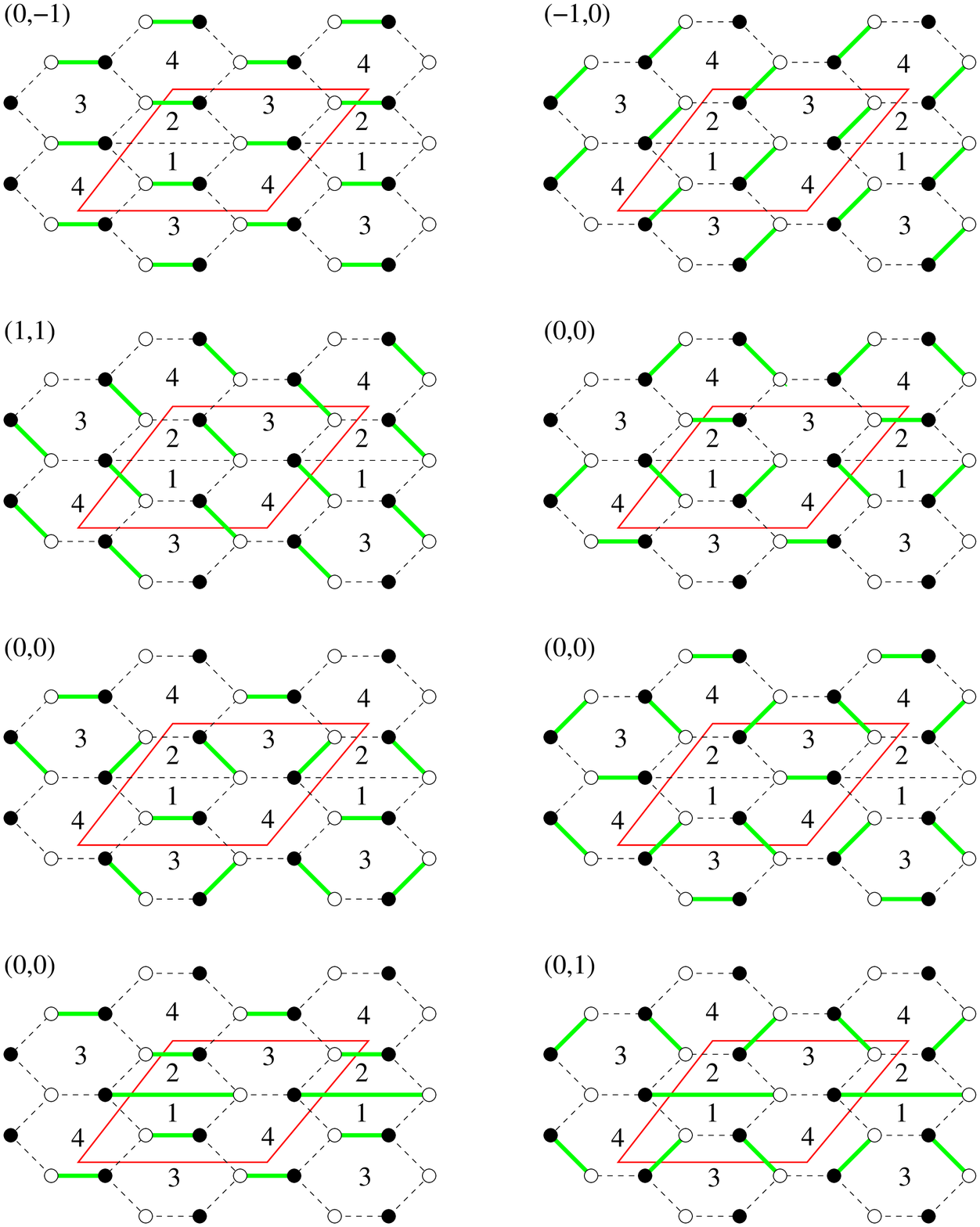}}
  \caption{The eight periodic perfect matchings of  ${\bf dP}_1$.
The green edges are contained in the matching.  The dashed lines are the edges left in the
tiling. The $(s,t)$ numbers are the corresponding points in the toric diagram (see Figure 8).}
  \label{dP1_mat}
\end{figure}

A {\bf perfect matching} is a subgraph of the tiling that contains all the nodes and each node
has valence one \cite{Kenyon:2003uj, Kenyon:2002a}. This means that a perfect matching is a set
of {\bf dimers} (edges in the brane tiling) that are separated, i.e.~they don't touch each
other; furthermore the dimers cover all the nodes. Therefore, we have altogether $V/2$ dimers in
each perfect matching, where $V$ denotes the number of nodes in the tiling. The eight perfect
matchings for ${\bf dP}_1$ are shown in \fref{dP1_mat}.

\begin{figure}[ht]
  \epsfxsize = 7cm
  \centerline{\epsfbox{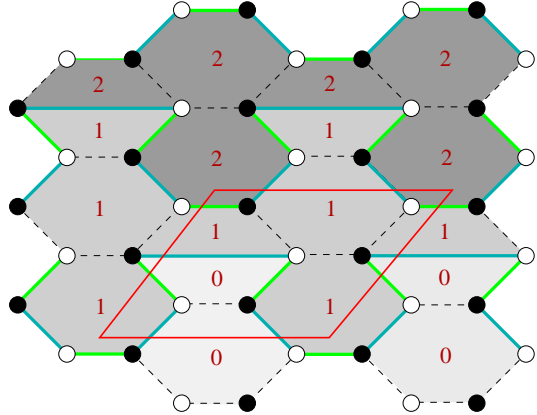}}
  \caption{The height function for the last perfect matching of ${\bf dP}_1$.
  The edges in the matching are colored blue. The green reference matching was chosen
  to be the $4^\textrm{th}$ perfect matching from Figure 9.
  The two matchings on top of each other result in horizontal loops where the height function
  increases by one.
  The monodromy of the height function is $(0,1)$ and there is a corresponding
  lattice point in the toric diagram in Figure 8.
  }
  \label{dP1_height}
\end{figure}

It can be easily checked that if we superimpose two perfect matchings $A$ and $B$ (denoted
$A+B$), then we obtain loops (and separate edges which we neglect). Fix a reference perfect
matching $R$. For each matching $A_i$ we can define an integer {\bf height function}. The loops
of $R+A_i$ can be regarded as ``contours''. Crossing a loop at an edge where the black node is
on the left hand side means a change in the height function by $\pm 1$. The sign depends on
whether the edge was part of the reference matching ($-1$) or that of $A_i$ ($+1$). An example
is shown in \fref{dP1_height} which is the height function for ${\bf dP}_1$ for the last
matching in \fref{dP1_mat}. The shading indicates the height. The contours are made of blue and
green edges that are contained in the last matching in the list and in the reference matching
(4th matching in the list), respectively.

The above defined height function is a well--defined function on the infinite periodic tiling
faces, but on the torus it has monodromy that is described by two integers:
$(s,t)$.\footnote{$(s, t)$ denotes the change in the height as we go along the two non--trivial
cycles of the torus of the brane tiling. This pair is also known as the {\bf slope of the height
function}.} For our example this pair was $(0,1)$. Such pairs are assigned to every perfect
matching with respect to a reference matching. These pairs are coordinates of points in the
toric diagram; in fact, the toric diagram is the (convex) set of all such points. The reference
matching has $(0,0)$ coordinates and the change in the reference matching merely translates the
toric diagram. An $SL(2,\Z)$ transformed fundamental cell results in an $SL(2,\Z)$ transformed
toric diagram. Perfect matchings that reside at an internal lattice point in the toric diagram
are called {\bf internal matchings}. The remaining matchings at the external points are the {\bf
external matchings} with corresponding Weil divisors as discussed in section
\ref{section_quivers}.

\vskip 0.5cm

\subsection{Zig--zag and rhombus paths}
\label{subsec_zigzags}

\begin{figure}[ht]
\begin{center}
  \epsfxsize = 7cm
  \centerline{\epsfbox{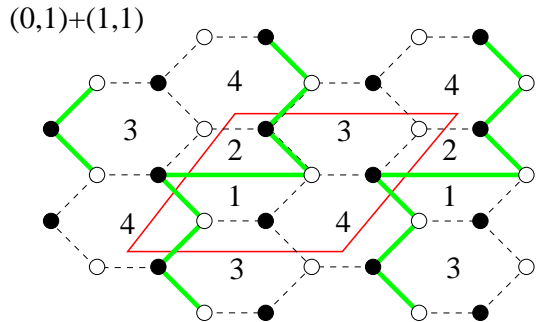}}
  \caption{Zig--zag path in the $\bf{dP}_1$ brane tiling.
The path is the superposition of the $(0,1)$ and $(1,1)$ neighboring external perfect
matchings.}
  \label{dP1zigzag}
\end{center}
\end{figure}

In \cite{Hanany:2005ss} a special path was defined which turns out to be also useful here. A
{\bf zig--zag path} is a path on the edges of the tiling which turns maximally left at a node,
then maximally right at the next node, then again left, and so on \cite{Kenyon:2002a}. An
example is presented in \fref{dP1zigzag}.

In \cite{Hanany:2005ss} it was an observation that these loops on the torus of the tiling are in
one--to--one correspondence with the edges of the toric diagram polygon. In fact, their homology
classes give the outward pointing normal vectors of the edges, the so--called external pq--legs.
This follows from the results of \textsection $7.1$ in \cite{Feng:2005gw}.

\begin{figure}[ht]
\begin{center}
  \epsfxsize = 11cm
  \centerline{\epsfbox{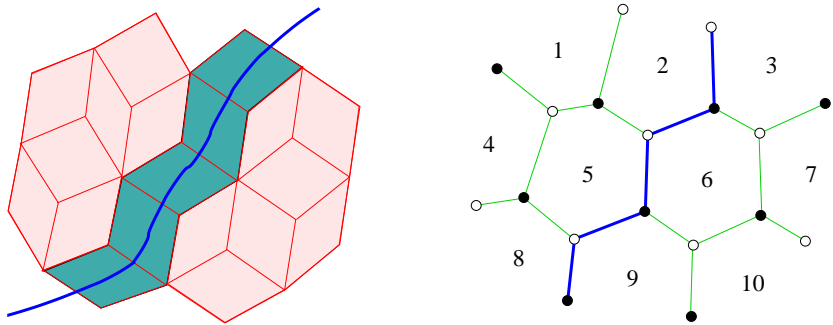}}
  \caption{(i) Rhombus path in the rhombus lattice.
(ii) Equivalent zig--zag path in the brane tiling. The blue line shows the rhombus loop
schematically. The edges which are crossed by the blue line in (i) are all parallel. Their
orientation can be described by an angle, the so--called rhombus loop angle.}
  \label{samplezigzag}
\end{center}
\end{figure}

As discussed at length in \cite{Hanany:2005ss}, one can associate to the brane tiling another
graph, which contains the edges that connect the face centers to the tiling nodes. We call this
graph the {\bf rhombus lattice}.\footnote{The faces of this new graph are indeed rhombi if the
original brane tiling is isoradially embedded in the plane. For details see
\cite{Hanany:2005ss}.} The zig--zags in the tiling are ``straight'' rhombus paths in the rhombus
lattice (\fref{samplezigzag}).

There are always two zig--zag paths going through each tiling edge. In the rhombus lattice this
translates to the fact that the bifundamentals arise from the intersection of two rhombus paths.
These paths have been analyzed in considerable detail in \cite{Feng:2005gw} where the zig--zags
were related to cycles that are wrapped by D6--branes in the mirror Calabi--Yau. Further
developments in brane tilings which will not be discussed here can be found in
\cite{Butti:2005vn, Butti:2005ps}.

 \clearpage

\section{Exceptional collections}
\label{section_excol}

Exceptional collections provide a powerful tool for deriving the low energy gauge theory
description of a stack of D--branes probing a Calabi--Yau singularity. Given a Calabi--Yau cone
$Y$, a stack of D--branes at the singularity will fragment into a set of fractional branes from
which the gauge theory is easily deduced. These fractional branes are best described as objects
in $D^ b(Y)$, the derived category of coherent sheaves on $Y$. Exceptional collections provide a
way of finding a good set of fractional branes and avoiding a direct confrontation with
$D^b(Y)$.\footnote{For earlier physics applications of exceptional collections to Landau
Ginzburg models, see \cite{Mayr:2000as, Tomasiello:2000ym, Govindarajan:2000vi, Hori:2000ck,
Zaslow:1994nk}.}

If $Y$ can be partially resolved by  blowing up a possibly singular complex surface $V$, instead
of looking for fractional branes on $Y$, we look for an exceptional collection of sheaves on
$V$.  There is then a simple procedure for converting this collection into a good set of
fractional branes  \cite{Herzog:2004qw, Aspinwall:2004vm}, and in fact the gauge theory can
often be deduced directly from the exceptional collection.

An {\bf exceptional collection} of sheaves ${\mathcal E} = (E_1, E_2, \ldots, E_n)$ is an
ordered set of sheaves which satisfy the following special properties:
\begin{enumerate}
\item Each $E_i$ is exceptional: $\Ext^q (E_i, E_i) = 0$ for $q>0$ and
$\Ext^0(E_i, E_i) = \Hom(E_i, E_i) = {\mathbb C}$.
\item $\Ext^q (E_i, E_j) =0$ for $i>j$ and $\forall q$.
\end{enumerate}
In these notes, we will be most interested in the case where the collection is {\bf strongly
exceptional}, in which case $\Ext^q (E_i, E_j) = 0$ for $i<j$ and $q>0$. For smooth toric
surfaces, the collection must be strong to generate a physical quiver gauge theory
\cite{Herzog:2004qw, Aspinwall:2004vm}, and the same is true for singular surfaces as
well.\footnote{For a recent gauge theory interpretation of more general exceptional collections,
see \cite{Wijnholt:2005mp}.}

For the most part, our sheaves can be thought of as line bundles, and line bundles are easy to
describe in a toric context.\footnote{ For singular surfaces when $D$ is not a Cartier divisor,
$\calo(D)$ is actually not a line bundle but only a reflexive sheaf. Nevertheless, for
simplicity, we will not emphasize this point further.} For each ray $v_r$ in the fan, there is a
toric Weil divisor $D_r$.  The line bundles can then be expressed as ${\mathcal O} \left( \sum_r
a_r D_r \right)$ for $a_r \in {\mathbb Z}$.
 One very special line bundle is the anti--canonical
bundle:
\be
{\mathcal O}(-K) = {\mathcal O} \left (\sum_r D_r \right) \ .
\ee
As we said earlier, the Calabi--Yau cone is the total space of the canonical bundle over our
surface. The fact that our fan defines a convex polygon means that $K$ is negative.


Given a strongly exceptional collection $\cale$, the  quiver gauge theory can be constructed
from the inverse collection $\cale^\vee$.  The members of $\cale^\vee$ are no longer sheaves but
objects in $D^b(V)$.  Lifting these objects to $Y$ yields the fractional branes. At the level of
D--brane charges, the inverse collection can be constructed from the Euler character on $V$,
$\chi(E_i, E_j^\vee) = \delta_{ij}$. As a set of objects in $D^b(V)$, $\cale^\vee$ is
constructed via a braiding operation called mutation described in detail in
\cite{Herzog:2004qw}.
The inverse collection is also exceptional although no longer strongly exceptional.  The Euler
character $\chi(E_i^\vee, E_j^\vee)$ can be interpreted as the number of arrows in the quiver
from node $i$ to node $j$ minus the number of arrows from node $j$ to node $i$
\cite{Herzog:2003dj, Aspinwall:2004vm}. This matrix is sometimes referred to as the
antisymmetric part of the adjacency matrix. More precisely, the Euler character tells us the net
number of $\Hom_{D^b(Y)}^1(E_i^\vee, E_j^\vee)$ maps in the Calabi--Yau between the fractional
branes.  For each of these maps, we have a massless open string which translates into a
bifundamental field in the quiver gauge theory.

It is often convenient to write down an intermediate quiver, the so--called Beilinson quiver,
which lives on $V$ instead of $Y$.  This quiver contains arrows corresponding only to the
negative entries of $\chi(E_i^\vee, E_j^\vee)$, or more precisely maps in $\Ext^1(E_i^\vee,
E_j^\vee)$. The Beilinson quiver algebra can be thought of as
\be
\oplus_{i,j} \Hom (E_i, E_j)  \ ,
\ee
but the quiver contains arrows only for the generators of this algebra which are encoded simply
in $\cale^\vee$. Because $V$ is compact, the Beilinson quiver contains no oriented loops.

\subsection{From Exceptional Collection to Periodic Quiver}

In this section we assume that we have a compact toric surface $V$ with positive anti--canonical
class and a strongly exceptional collection of line bundles ${\mathcal E}$ on $V$.  We would
like to construct from this data a periodic quiver.  In particular, we will write the Beilinson
quiver on a torus.

Any toric surface can be described by a fan by which we mean
 a collection of at least three vectors $v_r$, $r=1,\ldots, n$
on an integer lattice ${\mathbb Z}^2$.  That the surface is compact means that the polygon
defined by the endpoints of the vectors $v_r$ includes the origin.  That the anti--canonical
class of this surface is positive means that the polygon is convex.  (We would like to allow $V$
to have quotient singularities.)

One way of understanding $V$ is as a quotient of ${\mathbb C}^n$.  Given $n$ vectors in
${\mathbb Z}^2$, we expect that there will be $n-2$ linearly independent relations between the
$v_r$, which we write as
\be
\sum_r Q_{ar} v_r = 0 \label{charges}
\ee
where $a=1, \ldots, n-2$ and $Q_{ar} \in {\mathbb Z}$. Geometrically, we quotient
\be
\frac{{\mathbb C}^n - F_{\Delta}}{({\mathbb C}^*)^{n-2}}
\ee
where the action of the $({\mathbb C}^*)^{n-2}$ is given by the $Q_{ar}$.  The set $F_\Delta$ is
a small set of points inside ${\mathbb C}^n$ which we need to remove to have a well defined
quotient.

As an example, consider ${\mathbb P}^2$ for which the fan is $v_1 = (1,0)$, $v_2=(0,1)$, and
$v_3 = (-1,-1)$.  There is just one relation which we write as $Q = (1,1,1)$.  This quotient
construction is nothing but the usual equivalence relation of the homogenous coordinates on
${\mathbb P}^2$, namely $(X_1, X_2, X_3) \sim (\lambda X_1, \lambda X_2, \lambda X_3)$ for
$\lambda \in {\mathbb C}^*$. $F_\Delta$ is the origin $(0,0,0)$ of ${\mathbb C}^3$.

For arbitrary $V$, we can think of $X \in {\mathbb C}^n$ as generalized homogenous coordinates.
The $n-2$ equivalence relations (\ref{charges}) leave a two complex dimensional space which is
$V$ itself:
\be
(X_1, X_2, \ldots, X_n) \sim (\lambda^{Q_{a1}} X_1, \lambda^{Q_{a2}} X_2,\ldots,
\lambda^{Q_{an}} X_n) \ .
\ee

This two complex dimensional space $V$ is a fiber bundle $\pi: V \to B$ where $B$ is a real two
dimensional surface and the fibers are real, two dimensional tori.  More simply put, the fibers
are coordinatized by the phase angles of the complex coordinates on $V$. First, we characterize
this torus in greater detail.

Given the $n-2$ vectors $Q_{a}$ and using the standard inner product on ${\mathbb Z}^n$, we find
two additional vectors $q_1$ and $q_2$ such that $q_i \cdot Q_{a} = 0$ and $q_1$ and $q_2$ are
linearly independent. A canonical set of $q_i$ are the $v_r$ reinterpreted as two $n$
dimensional vectors rather than $n$ two dimensional vectors:  we could set $q_{1r} = v_{r,1}$
and $q_{2r} = v_{r,2}$. These $q_i$ can be used to measure relative positions on the real two
torus. Given the homogenous coordinates $(X_1, X_2, \ldots, X_n)$, we define the two torus
coordinates to be
\be
(\theta_1,\theta_2) = ( \sum_r q_{1r} \mbox{Arg} X_r, \sum_r q_{2r} \mbox{Arg} X_r) \ .
\ee
Notice that if we shift $X_r$ by $\lambda^{Q_{ar}}$, $(\theta_1,\theta_2)$ remains invariant
because $q_i \cdot Q_a = 0$.


Our D--branes are line bundles on $V$, and thus we can think of them as Euclidean D4--branes
filling all of $V$.  If we perform fiberwise T--duality twice on the two torus, we should find
D2--branes localized at points on the torus.  The open strings will then connect these points
together.  The periodic Beilinson quiver is nothing but this web of D2--branes and open strings.

We will characterize this web using the original line bundle (or D4--brane) description.  The
notation $\calo(D)$ indicates a D4--brane with a dissolved D2--brane; this dissolved D2--brane
has the same charges as a D2--brane wrapping the divisor $D \subset V$.  We can describe this
dissolved D2--brane as magnetic flux.  Because the line bundle is holomorphic, the field
strength components $F_{ij}=0=F_{\bar \imath \bar \jmath}$ vanish, and locally the field
strength takes the form
\be
F = i\partial_i \bar \partial_{\bar \jmath} (f + f^*) dy^i \wedge d{\bar y}^{\bar \jmath} \
\ee
where $A_j = -i \partial_j f$, $A_{\bar \jmath} = i \bar \partial_{\bar \jmath} f^*$ and $f$ is
some function of the coordinate patch. By a gauge choice, we may take the imaginary part of $f$
to vanish.

In a toric variety, the phase angle directions $\theta_i$ are isometries, and the field strength
$F$ describing the D2--brane should not depend on the $\theta_i$.  Because our variety is toric,
we can choose a complex structure such that $y^j = \ln r_j + i\theta_j = \rho_j + i\theta_j$. In
this coordinate system, the field strength becomes
\be
F = \left( \frac{\partial^2 f}{\partial \rho_i \partial \rho_j} + \frac{\partial^2 f}{\partial
\theta_i \partial \theta_j} \right) d\rho_i \wedge d\theta_j + \frac{\partial^2 f}{\partial
\theta_i \partial \rho_j} (d\rho_i \wedge d\rho_j + d\theta_i \wedge d\theta_j) \ .
\ee
In order for $F$ to be independent of $\theta_i$, $f$ must take a very special form.  In
particular, $f = g(r) + C_{i\bar \jmath} y^i {\bar y}^{\bar \jmath}$ where the second term leads
to a constant field strength.  We will assume this second term in $f$ vanishes in which case the
vector potential takes the very simple form
\be
A = \frac{ \partial f}{\partial \rho_i} d \theta_i \ . \label{Acanonical}
\ee



At this point, we fix a point $(r_1, r_2) \in B$ and look at the $T^2$ fiber, where we recognize
a Wilson line. Locally on the $T^2$, $A = w_j d\theta_j $ is pure gauge; $A =  i d \ln \Lambda$
where $\Lambda = \exp(-i w_j \theta_j)$.  However, globally, $\Lambda$ does not respect the
periodicity conditions.  We have a distinct set of Wilson lines for $0 \leq w_j < 1$, with
$(w_1, w_2) \sim (w_1 + n, w_2 + m)$ for $n$ and $m$ integers. This set of Wilson lines lives on
a dual torus we will call $\tilde T^2$.

Given a collection of line bundles, we can calculate the value of the Wilson line for each such
bundle and plot that point $(w_1, w_2)$ on our $\tilde T^2$ of length and height one.  This plot
gives us the nodes of the periodic Beilinson quiver.

The strings between the D4--branes come from the generators of the Beilinson quiver algebra and
as such are maps of the form $\Hom(E_i, E_j)$.  Since the branes are line bundles, we may write
$E_i = \calo(D)$, $E_j = \calo(D')$, and   $\Hom(E_i, E_j) = H^0(V, \calo(D'-D))$. We expect,
given a generating element in $\Hom(E_i, E_j)$, to find a corresponding string between
$\calo(D)$ and $\calo(D')$.  Moreover, $\calo(D)$ and $\calo(D')$ should be separated by a
vector on the torus given by the value of the Wilson line for $\calo(D'-D)$.

From the derived category point of view on $Y$, we know how to compute the masses of these open
strings \cite{Herzog:2004qw, Aspinwall:2004vm, Aspinwall:2004mb}, and the answer depends on
being able to understand instanton corrections as we move in the K\"ahler moduli space of $Y$.
From the point of view of the complex surface $V$ and the Wilson line discussion, our intuition
is that a string stretching between two of these D4--branes will have a mass proportional to the
distance between the corresponding points on $\tilde T^2$ \cite{Polchinski}.  As we change the
base point, the Wilson lines will all move around. Our naive expectation is that for massless
strings, there is a particular choice of base point for which the Wilson line corresponding to
$\calo(D'-D)$ vanishes.  It would be interesting to understand these masses better from the
Wilson line point of view.


\subsection{Line Bundles and Curvature Forms for Toric Surfaces}

In the previous section, we sketched a procedure for converting a set of line bundles on a toric
variety into a periodic quiver, but we did not explain why the construction would respect the
periodicity of the torus.  For example, take two linearly equivalent divisors $D$ and $D'$. The
corresponding line bundles $\calo(D)$ and $\calo(D')$ correspond to the same D--brane.  Why then
are the Wilson lines for $\calo(D)$ and $\calo(D')$ the same? In this section, we will attempt
to answer this question and elucidate the structure of the corresponding vector potentials.

Given a line bundle, $\calo(D)$, and a particular choice of K\"ahler metric on a toric variety,
one can construct an explicit coordinate dependent expression for a representative of $c_1(D)
\in H^2(V, {\mathbb Z})$.  These representatives were first worked out by \cite{Guillemin} (for
a readable and more recent account see \cite{Abreu}). This representative of $\calo(D)$ is
holomorphic, i.e. locally of the form $i \partial \bar \partial f$.  Also, it is independent of
the angular coordinates $\theta_i$ and so takes the form (\ref{Acanonical}) discussed
previously.

These representatives have a number of disadvantages. In most cases, these representatives do
 not satisfy the remaining
equation of motion $g^{i\bar \jmath}F_{i\bar \jmath} = \mu$. Here, $\mu$ is a constant often
called the slope. Moreover, they depend on a particular canonical choice of K\"ahler metric
which is usually not the one of physical interest.  Typically, we would be more interested in a
metric which is compatible with a Ricci flat metric on the cone over $V$.\footnote{ It may be
that the metric compatible with a Ricci flat metric on the cone is not K\"ahler.  For example,
the metric on $\mathbf{dP}_1$ compatible with the $Y^{2,1}$ Sasaki--Einstein metric is not
K\"ahler \cite{Martelli:2004wu}.} Despite these disadvantages, we use these explicit
representatives for they form a useful beginning from which to argue more general results.

We have thus far been working with complex coordinates $\rho + i \theta$, but these
representatives are most easily expressed in symplectic coordinates on $V$, $x + i\theta$. The
phase angles $\theta_i$ remain the same in both the complex and symplectic system.  For the $x$,
we define a polytope
\be
\Delta = \{x \in {\mathbb R}^2 : \langle x, v_r \rangle \geq -1 \, \,  \forall r \} \ .
\ee
The symplectic form is then $\omega = \sum_i dx_i \wedge d\theta_i$.

In these symplectic coordinates, the K\"ahler metric and complex structure depend on a potential
function $g(x)$.  Define
\be
g_{ij} = \frac{\partial^2 g(x)}{\partial x_i \partial x_j} \ .
\ee
The line element becomes
\be
ds^2 = g_{ij} dx_i dx_j + g^{ij} d\theta_i d\theta_j
\ee
where $g^{ij}$ is the inverse of $g_{ij}$ and summation on the indices is implied. The
symplectic coordinates are related to the complex ones by a Legendre transformation, $\rho =
\partial g / \partial x$.

The representatives of $H^2(V, {\mathbb Z})$ depend on a particular choice of $g$,
\be
g_{can} = \frac{1}{2} \sum_r \ell_r \log \ell_r \ ,
\ee
where we have defined
\be
\ell_r =  \langle x, v_r \rangle +1 \ .
\ee
In the case of projective space, this metric is physically interesting: it's Einstein and is
thus compatible with a Ricci flat metric on the cone over $V$.  In general $g_{can}$ will
produce a metric which is physically uninteresting albeit simple.  A general K\"ahler metric is
related to $g_{can}$ in a smooth way:
\be
g = g_{can} + h
\ee
where $h$ is a smooth function on $\Delta$.

We have seen already that a holomorphic vector bundle has a curvature form which may be written
as $2i \partial \bar \partial f(\rho)$ for some locally defined function of $f$.  In symplectic
coordinates, this two--form becomes
\be
2i \partial \bar \partial f = \sum_{j,k} \frac{\partial}{\partial x_j} \left(g^{kl}
\frac{\partial f}{\partial x_l} \right) dx_j \wedge d \theta_k \ . \label{Fgeneral}
\ee
For the canonical choice of metric, we take the vector potential  corresponding to $\calo(D_r)$
to be
\be
A_r = \frac{1}{2} (g_{can})^{kl} \frac{\partial \log \ell_r}{\partial x_l} d\theta_k \ .
\label{vecform}
\ee
This $A_r$ yields a curvature two--form which represents the class $c_1(D_r)$ but is in general
not harmonic. Note that $A_r$ is only well defined away from the side $\ell_r = 0$.

Using (\ref{vecform}), we will prove a result about the $A_r$ and then argue that the same
result must hold more generally for non--canonical metrics and $A_r$ which do satisfy the
equations of motion.  The result is that
\be
\sum_r v_{r,i} A_r = d\theta_i \label{cangauge}
\ee
or in other words, this particular combination of the $A_r$ is pure gauge.  The result follows
simply from noting that
\be
(g_{can})_{ij} = \sum_r \frac{v_{r,i} v_{r,j}}{2 \ell_r} \ . \label{gij}
\ee

More generally, because every divisor $D = \sum_r a_r D_r$ can be expressed as a sum of
primitive Weil divisors, we expect there to be a basis of primitive vector potentials $A_r$, $r
= 1 ,\ldots, n$ such that $A_D = \sum_r a_r A_r$.  We have now chosen the $A_r$ to satisfy the
equations of motion, but they should be related to the canonical $A_r$ in a smooth way. We say
two divisors $D$ and $D'$ are linearly equivalent when they have the same $Q$ charges, $\sum_r
Q_{ar} (a_r - a_r') = 0$.  All such linear equivalence relations are generated by the $q_i$.  If
$D$ and $D'$ are linearly equivalent, then $\calo(D-D') \sim \calo$. But $\calo$ corresponds to
a single D4--brane with no dissolved D2--brane charge.  The associated field strength must
vanish, and it must be that
\be
\sum_r q_{ir} A_r \label{gaugereq}
\ee
 is pure gauge for $i=1$ and 2.

We can deduce more from the statement that (\ref{gaugereq}) is pure gauge.  A gauge
transformation $A \to A + i d \ln \Lambda$ must respect the periodicity of the torus.  Since the
$A_r$ take the form $f(r) d\theta$, the gauge transformation $\ln \Lambda$ which annihilates
(\ref{gaugereq}) must depend only linearly on $\theta$ and not at all on $x$.  The only choice
is $\Lambda = \exp(i n \theta)$, from which we conclude that
\be
\sum_r q_{ir} A_r = n_{i1} d\theta_1 + n_{i2} d \theta_2
\ee
for integers $n_{ij}$. The $v_r$ and our Wilson line torus are only defined up to an
$SL_2({\mathbb Z})$ transformation so we choose
\be
\sum_r q_{1r} A_r = d\theta_1 \; ; \; \; \; \sum_r q_{2r} A_r = d\theta_2 \ ,
\ee
recovering the canonical result (\ref{cangauge}) in a more general context. This reasoning
answers the question posed earlier about why for linearly equivalent $D$ and $D'$, $\calo(D)$
and $\calo(D')$ give the same Wilson line.

Before moving on, we study the vanishing of the generating set $A_r$ because of a possible
relation to massless open strings.
 We wish to show that the $A_r$ will vanish at corners of $\Delta$
where $A_r$ is well defined.
 For this demonstration, we rely on a result of Abreu \cite{Abreu} that
 \be
 \det(g_{ij}) = \left[ \delta(x) \prod_{r=1}^n \ell_r(x) \right]^{-1} \ ,
 \ee
 where $\delta$ is a smooth function on $\Delta$.  Since we are on a surface, at a corner of
 $\Delta$, the determinant of $g^{ij}$
 involves a double zero, and it is straightforward to show that $g^{ij}$
 must vanish.  Since $g^{ij}$ vanishes, from (\ref{Fgeneral}) we see that $A_r$ will vanish
 as well unless the corner is associated with the vanishing of $\ell_r$.

\subsection{Bundles on ${\mathbb P}^2$ }

To illustrate these ideas concretely, we present them for ${\mathbb P}^2$. There are three Weil
divisors $D_1$, $D_2$, and $D_3$ on ${\mathbb P}^2$ corresponding to the three rays of the fan
$v_1 = (1,0)$, $v_2 = (0,1)$, and $v_3 = (-1,-1)$. From (\ref{vecform}), the vector potentials
for the corresponding three line bundles, which in this case satisfy the equations of motion,
are
\be
A_1 = -\frac{1}{3} (x_1-2) d\theta_1 - \frac{1}{3}(1+x_2)d\theta_2 \ , \label{firstp2ai}
\ee
\be
A_2 = -\frac{1}{3}(1+x_1) d\theta_1 - \frac{1}{3} (x_2-2) d\theta_2 \ ,
\ee
\be
A_3 = -\frac{1}{3} (1+x_1) d\theta_1 - \frac{1}{3} (1+x_2) d\theta_2 \ , \label{p2ai}
\ee
where the $x_i$ lie inside the triangle defined by $x_1 > -1$, $x_2 > -1$ and $x_1+x_2<1$. These
$A_r$ are all gauge equivalent to each other, which is expected since the corresponding divisors
are all linearly equivalent.  The gauge transformation takes the form $A \to A + d \lambda$
where $\lambda = n_1 \theta_1 + n_2 \theta_2$ and $n_i$ is an integer.  The Wilson line
corresponding to the $A_r$ will not change because the gauge transformation respects the
periodicity of this square torus of height and length one.  Thus we see that $\calo(D_1)$,
$\calo(D_2)$ and $\calo(D_3)$ appear as the same point on $\tilde T^2$.  Indeed, for any line
bundle of the form $\calo(a D_1 + b D_2 + c D_3)$, the point on the torus will depend only on
$a+b+c$. Any line bundle of the form $\calo(a D_1 + b D_2 + c D_3)$ can equivalently be written
as $\calo(a+b+c)$.

We can take the vector potential corresponding to $\calo(n)$ to be
\be
-\frac{n}{3}(1+x_1) d\theta_1 - \frac{n}{3}(1+x_2) d\theta_2 \
\ee
Thus, given the exceptional collection $\calo, \calo(1), \calo(2)$, we should plot points at
$p_1 = (0,0)$, $p_2 = (-1-x_1, -1-x_2)/3$ and $p_3 = 2(-1-x_1, -1-x_2)/3$ or their translates on
$\tilde T^2$.  These three points correspond to the D--branes.

\begin{figure}[ht]
\epsfxsize = 2in \centerline{\epsfbox{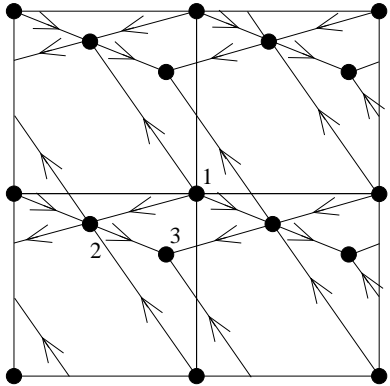}} \caption{Four unit cells of the ${\mathbb
P}^2$ periodic quiver for basepoint $(x_1, x_2) = (3/4, -1/2)$.} \label{p2periodic}
\end{figure}

To connect these three D--branes with open strings, we return to the $A_i$
(\ref{firstp2ai})--(\ref{p2ai}). Between $\calo$ and $\calo(1)$ or between $\calo(1)$ and
$\calo(2)$, there are three possible paths corresponding to $D_1$, $D_2$, and $D_3$.  The path
corresponding to $D_i$ is defined by the Wilson line associated to $A_i$.  Instead of thinking
of the Wilson line as a point on the torus, we now think of it as a vector that joins two
points.  The resulting Beilinson quiver for ${\mathbb P}^2$ is shown in Figure \ref{p2periodic}.
We do not need to draw in additional arrows corresponding to maps between $\calo$ and
$\calo(2)$. All the requisite maps can be formed by joining together the arrows already drawn.

%
These vectors corresponding to the $D_i$ shrink to zero size at special base points on the
polytope $\Delta$.  In particular, the string corresponding to $D_1$ shrinks to zero at
$(2,-1)$, $D_2$ shrinks to zero at $(-1,2)$, and $D_3$ shrinks to zero at $(-1,-1)$.

One startling feature of this Beilinson quiver is that the arrows will never cross, no matter
what our choice of basepoint $(x_1, x_2)$.  As the $(x_1, x_2)$ moves to the boundaries of
$\Delta$, arrows may become parallel and the three points may touch, but the arrows never cross.

\subsection{Constructing the Quiver in General}
\label{sec:construction}

Given a set of generating field strengths for the $\calo(D_r)$, we can construct a family of
periodic quivers from an exceptional collection.  A particular quiver in the family will depend
on the choice of basepoint $(x_1, x_2) \in \Delta$. If the metric is of physical interest,
e.g.~it lifts to a Ricci flat metric on the cone and provides a starting point for AdS/CFT
constructions, and the field strengths satisfy the equations of motion, we expect this periodic
quiver to be the quiver of physical interest. Thus, the quiver we described for ${\mathbb P}^2$
should be the ``correct'' quiver. Unfortunately, we in general do not have explicit expressions
for the metric and the field strengths, only the canonical representatives detailed above.

In the absence of physical data, we will work with the canonical metric and hope that the
resulting quiver is topologically if not geometrically accurate. Because we only expect
topological data, we will fix a particularly convenient choice of basepoint in $\Delta$: $(x_1,
x_2) = (0,0)$. In this case, the vector potential becomes
\be
A_r = \frac{1}{2} g^{kl} v_{r,l} d\theta_k \ .
\ee
From this vector potential, we see that a general line bundle of the form $\calo(\sum_r a_r
D_r)$ will be plotted on the torus with coordinates
\be
\left( \sum_r q_{1r} a_r, \sum_r q_{2r} a_r  \right) \ ,
\ee
where
\be
q_{ir} = \frac{1}{2} g^{il} v_{r,l} \ .
\ee
These two $q_{ir}$ are orthogonal to the $Q_a$ and are in fact the same as the $q_i$ discussed
previously. Because $g^{kl}$ is complicated and we are after only topological information, let
us rescale the $q_{ir}$ and the associated torus by a $g_{kl} \in GL_2({\mathbb R})$
transformation, choosing $q_{jr} = v_{r,j}$ as before.

The procedure for constructing the quiver is very simple. Given a strongly exceptional
collection of line bundles $\cale = (E_1, E_2, \ldots, E_n)$, take $E_j = {\mathcal O}(\sum_r
a_r D_r)$ and $E_{k} = {\mathcal O}( \sum_r b_r D_r)$. The homomorphisms from $E_j$ to $E_{k}$
are generated by the global sections of ${\mathcal O}(\sum_r (b_r - a_r) D_r)$.  Start with the
monomial
\be
\prod_r X_r^{b_r-a_r} \ .
\ee
This monomial has charges $\sum_r Q_{ar} (b_r - a_r)$. To be a global section, $b_r - a_r \geq
0$ for all $r$ (or there will be a pole).  However, there may be more than one such monomial
with this charge.  Construct all such monomials.  Call the set of such monomials $M_{jk}$. For
each $m \in M_{jk}$, where $m = \prod_r X_r^{c_r}$, we compute
\be
(\phi_1, \phi_2) = \left( \sum_i q_{1r} c_r, \sum_i q_{2r} c_r \right)
\ee
This vector $(\phi_1,\phi_2)$ is the relative position of nodes $j$ and $k$ on $\tilde T^2$.
Fixing the position of $E_1$, we now have specified the location of all the nodes of the quiver.

Instead of a $\tilde T^2$ of length and height one as before, because of the rescaling, the
period vectors of this torus are the $q_i$.  If we take two points of the quiver separated by $a
q_1 + b q_2$, in the language of line bundles, we have $\calo(D)$ and $\calo(D + \sum_r (a
q_{1r} +b q_{2r}) D_r)$.  However, since the $q_i$ are orthogonal to the $Q_i$, $D$ and
$D+\sum_r(a q_{1r} + b q_{2r} ) D_r$ have the same $Q$ charges and are linearly equivalent as
divisors. In other words, these two points are the same.

Starting with the set $M_{k,k+1}$, we draw an arrow from node $k$ to node $k+1$ for each $m\in
M_{k,k+1}$. We repeat this procedure for line bundles of the form $E_k$ and $E_{k+2}$. There is
an additional complication now.  It may happen that the monomial $m = m_1 m_2$ where $m_1$ joins
nodes $E_k$ with $E_{k+1}$ and $m_2$ joins nodes $E_{k+1}$ and $E_{k+2}$.  If such is the case,
then we do not add an arrow corresponding to $m$. The entries of $\chi(E_i^\vee, E_j^\vee)$ let
us know how many arrows we should be writing down. Recursively, we consider $E_k$ and $E_{k+i}$
and continue until all the arrows in the Beilinson quiver are drawn.

\begin{figure}[ht]
  \epsfxsize = 3in
  \centerline{\epsfbox{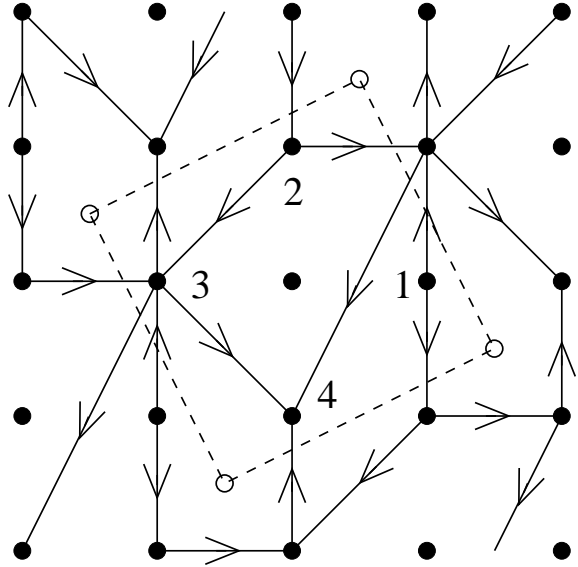}}
\caption{The periodic Beilinson quiver for $\bf{dP}_1$ with fundamental cell.}
\label{dP1Beilinson}
\end{figure}

Take $\bf{dP}_1$ to illustrate these ideas. A fan is $v_1 = (0,1)$, $v_2=(1,1)$, $v_3 = (0,-1)$,
and $v_4 = (-1,0)$ from which we choose
\be
q = \left(
\begin{array}{cccc}
0 & -1 & 0 & 1 \\
-1 & -1 & 1 & 0
\end{array}
\right) \ .
\ee
An exceptional collection on $\bf{dP_1}$ is $\calo, \calo(D_1), \calo(D_4 + D_1),
\calo(D_4+D_1+D_3)$. Using the procedure described above, we find the Beilinson quiver, figure
\ref{dP1Beilinson}.

For example, consider the paths between $\calo(D_4+D_1)$ and $\calo(D_4+D_1+D_3)$. We look for
all monomials with the $Q$ charges of $D_3$, in other words $x_3$, $x_1 x_4$, and $x_1 x_2$.
These three monomials have torus charges $q$, $(0,1)$, $(1,-1)$, and $(-1,-2)$ respectively.  On
our torus, node 4 is indeed at relative positions $(0,1)$, $(1,-1)$, and $(-1,-2)$ to node 3
with corresponding arrows drawn in.

\subsection{Vanishing Euler Character}

We can argue that the Euler character of the torus (to be distinguished from the Euler character
of the exceptional collection) must vanish and so the most obvious obstruction to writing the
quiver on a torus is eliminated.  (Of course, we don't have an arbitrary collection of lines,
vertices, and faces, but have instead completely specified the connectivity, and it remains
unclear that the pattern of connectivity will be compatible with a torus structure.)  Given
exceptional collections $\cale$ and $\cale^\vee$, in terms of charges, we can decompose any
sheaf $F$ into the $E_j^\vee$ or the $E_j$:
\be
\ch(F) = \sum_j \chi(E_j, F) \ch(E_j^\vee) \; ; \; \; \; \ch(F) = \sum_j \chi(F, E_j^\vee)
\ch(E_j)  \ .
\ee
We are interested in quivers that come from a stack of D3--branes, which look like a point in
$V$. Thus, for a skyscraper sheaf
\be
\ch({\mathcal O}_{pt}) = \sum_j \chi(E_j, {\mathcal O}_{pt}) \ch(E_j^\vee) = \sum_{i,j}
\chi(E_j, {\mathcal O}_{pt}) \chi(E_j^\vee, E_i^\vee) \ch(E_i) \ .
\ee
The rank component of the chern class of a skyscraper sheaf vanishes, and $\chi(E_i, {\mathcal
O}_{pt}) = \rk(E_i)$. Thus,
\be
0 = \sum_{i,j} \rk(E_i) \rk(E_j) \chi(E_i^\vee, E_j^\vee) \ .
\ee
For these toric exceptional collections, we find exceptional collections of line bundles where
the ranks are all one.  Thus, the sum over the entries of the Euler character must vanish. But
this sum has a different interpretation.  The sum over the diagonal entries is the number of
gauge groups.  The sum over the negative entries is the number of arrows in the Beilinson
quiver, and the sum over the off--diagonal positive entries is the number of relations:
\be
\sum_{i,j} \chi(E_i^\vee, E_j^\vee) = \mbox{gauge groups} - \mbox{arrows} + \mbox{relations} \ .
\ee
Now for these toric quivers, we know that each relation corresponds to two superpotential terms.
Moreover, when we lift to the Calabi--Yau quiver, each relation also becomes an additional
arrow.  Thus, for the Calabi--Yau quiver
\be
\mbox{gauge groups} - \mbox{arrows} + \mbox{superpotential terms} = 0
\ee
which is exactly the condition that the Euler character of the torus vanish because for each
gauge group we have a node, for each arrow an edge, and each superpotential term a face in the
quiver.\footnote{We would like to thank Aaron Bergman for this observation relating
$\chi(E_i^\vee, E_j^\vee)$ to the Euler character of the torus.} Moving back to the Beilinson
quiver now consists of removing a set of arrows, which cannot change the Euler character of the
graph. This demonstration of vanishing Euler character is complementary to but distinct from a
similar observation in \cite{Franco:2005rj} where the authors use R--charge constraints to prove
that the Euler character of the brane tiling vanishes.

\section{Compatibility}
\label{section_compa}

Having established that one can derive periodic quivers from exceptional collections, we now
study the possibility of generating such collections by means of brane tilings. In this section
we define a map that assigns line bundles to paths in the quiver. This map can be used to
compute an exceptional collection on a complex surface that shrinks to zero size at the
singularity. The exceptionality can be checked on a case--by--case basis. Given these bundles,
one can reconstruct the quiver based on mathematically rigorous procedures \cite{Cachazo:2001sg,
Wijnholt:2002qz, Aspinwall:2004bs, Aspinwall:2005ur, Herzog:2004qw, Aspinwall:2004vm,
Herzog:2005sy, Herzog:2003zc}. By reinterpreting paths and perfect matchings in the tiling
language, we explicitly prove that this construction gives back our original quiver.

\subsection{Beilinson quivers and internal matchings}
\label{subsec_beil}

For the exceptional collection technique to be useful when applied to toric Calabi--Yau
manifolds, we need the toric diagram to contain at least one internal point. This restriction
means that our manifold can be partially resolved by blowing up a 4--cycle. Let us consider the
tiling for this Calabi--Yau which can be most efficiently constructed by the Fast Inverse
Algorithm \cite{Hanany:2005ss, Feng:2005gw}. Let us also fix a reference internal matching
$PM_0$ that resides at one of the internal points of the toric diagram. We can set the origin at
this point.

If we remove those bifundamentals from the quiver that are contained in $PM_0$, then we obtain
another smaller quiver. We will show that this subquiver contains no oriented loops and
therefore has the right properties to be a Beilinson quiver for the relevant 4--cycle.\footnote{
We would like to thank Robert Karp for discussion about this point.} For an example see
\fref{dPbeil}. This Beilinson quiver is generated by deleting the bifundamentals that are
contained in the $4^\textrm{th}$ perfect matching of \fref{dP1_mat}.  Recall that the Beilinson
quiver was defined at the beginning of Section \ref{section_excol} from an exceptional
collection. Here, we define an intermediate notion \bdefn We define a {\bf pre--Beilinson
quiver} to be a connected subquiver of the gauge theory quiver that contains no oriented loops
and all the nodes of the original. \edefn

Let us summarize some additional terminology we use in the following.
\bdefn An {\bf oriented path} is a path in the quiver that respects the direction of the arrows.
\label{orientedpath} \edefn
\bdefn Paths in the quiver that also exist in a Beilinson (or pre--Beilinson) quiver are called
{\bf allowed paths}. \label{allowedpath} \edefn
We say that a path crosses an edge in the tiling if the path contains the corresponding arrow in
the quiver.  Paths that exist in the Beilinson quiver will not intersect the edges of $PM_0$. It
is easy to see that F--terms transform allowed paths to allowed paths. Closed paths may wind
around the tiling torus, and the winding can be characterized by the homology class of the loop
($p,q$). The ($0,0$) loops are called {\bf trivial loops}. By definition, the {\bf length of an
oriented path} is the R--charge of the corresponding operator. Paths can be related by F--term
transformations, but these transformations will not change the total R--charge associated to a
path. The height functions of the external matchings with respect to $PM_0$ are called {\bf
height coordinates}.

\blemma{In a consistent tiling, an internal perfect matching determines a pre--Beilinson quiver
by removing those bifundamentals from the quiver that are contained in the matching.}
\elemma
\bproof

Removing bifundamentals from the gauge theory quiver that are contained in $PM_0$ does not
remove nodes and does not create disconnected pieces.  The nontrivial part of the proof involves
the oriented loops.

\vskip 0.3cm

\begin{figure}[ht]
  \epsfxsize = 6cm
  \centerline{\epsfbox{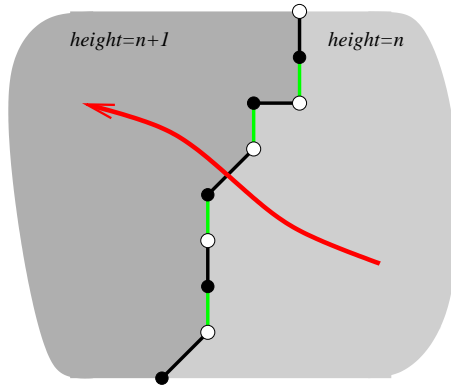}}
  \caption{Allowed face paths (i.e.~paths in the Beilinson quiver) go always uphill.
  The height function increases by one at the line constituted of the black perfect matching
  and the green reference matching. The red path cannot cross the green edges (they
  are not in the Beilinson quiver). Hence when crossing the contour line, the red path
  has to cross a black edge.
  Crossing the black edge increases the value of the height function.}
  \label{levelcross}
\end{figure}

(i) First we show that trivial allowed loops cannot exist. Such a loop would contain at least
one edge $e$. By crossing this edge in the tiling, some of the height functions would increase
by one. The increase happens exactly when the corresponding perfect matchings contain $e$.
Allowed paths will never go ``downhill'' on the graph of any height function, because then they
would have to cross an edge in $PM_0$ which is not allowed (the edge is not present in the
pre--Beilinson quiver). See \fref{levelcross} for the schematic picture. The increase of the
height function is ``irreversible'', i.e.~the function is monotone along an allowed path; hence
we have arrived at a contradiction.

For this argument to hold one has to show that $e$ is contained in at least one perfect
matching. We can suppose this, since otherwise we can omit this edge from the tiling and still
get the same toric diagram which questions the consistency of the original tiling.

\begin{figure}[ht]
  \epsfxsize = 3.5cm
  \centerline{\epsfbox{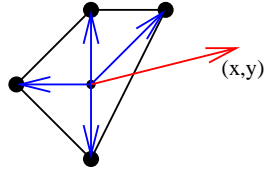}}
  \caption{Gradient vectors in the toric diagram. The coordinates of the blue $(s_i,t_i)$ vectors give the monodromy
  of the height function of the perfect matching sitting at their endpoints. The red $(x,y)$ arrow is
  the gradient vector of the hypothetical nontrivial loop.}
  \label{td_stxy}
\end{figure}

\vskip 0.3cm (ii) We also need to show that there are no non--trivial loops in the
pre--Beilinson quiver. These non--trivial loops wrap the torus cycles. Suppose that there exists
such a loop. This oriented loop is a face path on the brane tiling with homology class $(x,y)\in
\IZ^2$ as in \fref{td_stxy}. Let us take an arbitrary external matching $PM_i$ at $(s_i,t_i)$.
We can compute the height function assigned to this matching with respect to $PM_0$.

The height function should not decrease along the path. As an immediate consequence, the scalar
product $(s_i,t_i)\cdot (x,y)$ must be nonnegative. On the other hand, the set of vectors
$\{(s_i, t_i) \}$ span the whole 2d space with positive coefficients, and thus at least one of
these vectors has negative scalar product with $(x,y)$. This is a contradiction; therefore the
pre--Beilinson quiver doesn't contain non--trivial loops. \eproof

\subsection{Line bundles from tiling: The $\Psi$--map}
\label{subsec_psi}

In the last section we saw that a candidate Beilinson quiver could be created from an {\it
internal perfect matching}. In this section we continue by defining a map $\Psi$ that assigns a
divisor to an allowed path by using {\it external perfect matchings}. We conjecture that these
divisors give exceptional collections of line bundles which we will use to reconstruct the
Beilinson quiver.

\begin{figure}[ht]
  \epsfxsize = 4cm
  \centerline{\epsfbox{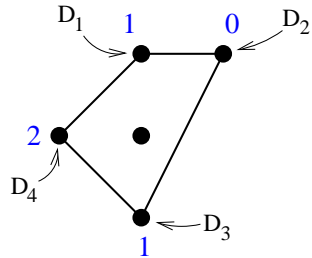}}
  \caption{An integer function over the external nodes determines a divisor and therefore
a sheaf of sections of the corresponding line bundle. The numbers in the figure denote
${\mathcal O}(D_1+D_3+2D_4)$.}
  \label{dP1_dev}
\end{figure}

A Weil divisor can be represented by an integer function over the external vertices of the toric
diagram polygon (see \fref{dP1_dev}). We call two such integer functions equivalent if they
differ by a linear function $f(x,y)=xm+yn$ which defines a principal divisor. (Here $x$ and $y$
are coordinates on the plane of the polygon.)

\begin{figure}[ht]
  \epsfxsize = 13cm
  \centerline{\epsfbox{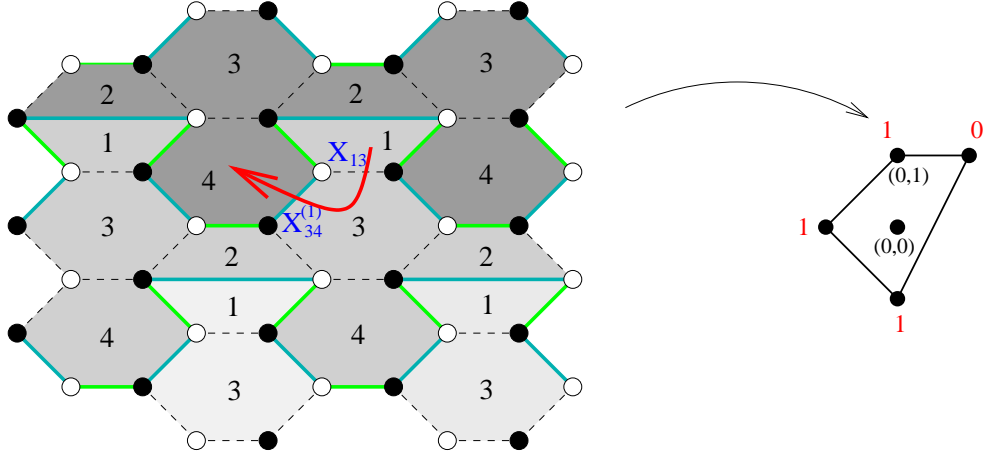}}
  \caption{The $\Psi$--map.}
  \label{dP1_psi}
\end{figure}

Let us fix an arbitrary oriented path $P$. Then, $\Psi(P)$ gives a divisor, i.e.~an integer
function over the external nodes. We define this map by using the matchings of the tiling. For
each external node $v_r$, there is a corresponding unique perfect matching\footnote{We assume
that the tiling is consistent and there are no ``external multiplicities'', i.e.~there is a
unique perfect matching corresponding to each external node of the toric diagram.} $PM_r$. We
assign to the divisor $D_r$ the integer $\Psi_r (P)$ that is the number of edges in $PM_r$ which
are crossed by the path $P$. In \fref{dP1_psi} we see an example.

The left hand side shows the brane tiling for ${\bf dP}_1$. The red path $P$ crosses two edges;
hence it labels the operator $X_{13} \cdot X_{34}^{(1)}$. There is a corresponding oriented $1
\rightarrow 3 \rightarrow 4$ path in the quiver as in Figure \ref{quiver_dp1}. We have chosen
the $4$th matching from \fref{dP1_mat} as the green reference matching. To show how to compute
$\Psi_8 (P)$, we have drawn the $8$th matching of Figure \ref{dP1_mat} (in blue). The shading of
the faces indicates the height function of this matching that has $(0,1)$ monodromy.
The red path crosses one blue edge in the matching (namely $X_{34}^{(1)}$); hence $\Psi_8 (P) =
1$. One can compute the other integer ``intersection numbers'' with the help of the other
external perfect matchings. The resulting numbers are indicated in red. These numbers define a
Weil divisor on the base of the threefold. The numbers can also be interpreted as the increase
in the height coordinates as we go along the path $P$. If the path is an allowed path
(Definition \ref{allowedpath}) starting at face $A$ and ending at $B$, then $\Psi_r$ is simply
the $h_r (B) - h_r (A)$ difference in the height function that corresponds to the
$r^{\textrm{th}}$ external node. $\Psi$ is a well--defined function on the paths of the quiver.
In fact, it does not depend on the choice of the reference perfect matching (modulo linear
equivalence).

The $\Psi$--map can be extended to unoriented paths, i.e.~paths that do not respect the arrow
direction in the quiver.  When crossing an edge in $PM_i$ in the reverse direction, we subtract
one instead of adding one in computing $\Psi_r(P)$.

Let $C_i$ denote the Abelian group of chains in the periodic quiver. Here the quiver is
understood as a discretization of the 2--torus. This is the free group generated by the edges in
the quiver with integer coefficients. The elements of $C_1$ take the following form
\be
  P = \sum_i c_i X_i \qquad (c_i \in \IZ)
\ee
where $X_i$ denotes the $i^\textrm{th}$ edge. We denote the cycles in $C_i$ by $Z_i$ and the
boundaries by $B_i$.  Elements of $B_1$ are built out of trivial loops. $\Psi$ can be extended
in a straightforward way to be defined on $C_1$
\be
  \Psi_r = \sum_j c_{p_j}
\ee
where $\{ p_j \}$ is the list of edges in the $r^\textrm{th}$ external matching. In the
following, we will study the properties of this extended $\Psi$--map.

For an elementary loop around a node in the tiling, the image of $\Psi$ is a {\bf constant
function} (the anticanonical class $K$). Since all the perfect matchings cover this node, each
matching is intersected by the loop precisely once; hence $\Psi_r=1$ for all $r$. This coincides
with the observations made in \cite{Franco:2005sm}. In fact, one can easily prove that the
entire $B_1$ subgroup is mapped to constant functions.

Gauge invariant mesonic operators can be constructed from arbitrary oriented
loops\footnote{Related work on mesonic operators was recently done in \cite{Kihara:2005nt,
Oota:2005mr}.}. These are the elements of $Z_1$. For these loops $\Psi$ assigns {\bf
non--negative affine functions} on the toric diagram parametrized by three integers. These
functions are points in the dual cone. This is being investigated in \cite{hanany_sparks}.

We will now use $\Psi$ to compute a {\bf collection of line bundles}. We choose an internal
reference matching which determines a Beilinson quiver and therefore an ordering of the faces in
the tiling. Without losing generality, we relabel the groups such that there are no arrows from
node $i$ to $j$ if $i>j$.

\begin{figure}[ht]
  \epsfxsize = 6cm
  \centerline{\epsfbox{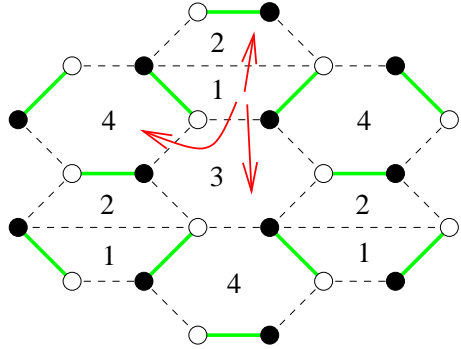}}
  \caption{The reference paths are allowed paths to each face. They start from face $1$ and
  don't cross the
edges of the green internal matching; hence they are paths in the Beilinson quiver.}
  \label{dP1_paths}
\end{figure}

\begin{figure}[ht]
  \epsfxsize = 8cm
  \centerline{\epsfbox{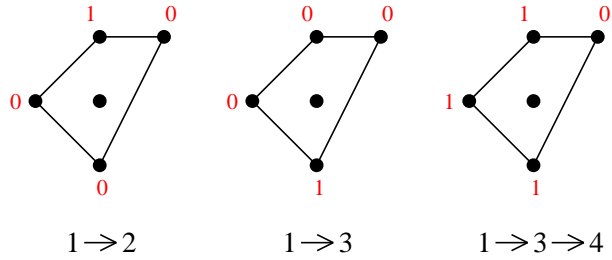}}
  \caption{The three divisors computed from the paths to the faces.}
  \label{dP1_divs}
\end{figure}

Let us fix an allowed path $P_i$ for each face in the tiling (for  ${\bf dP}_1$ see
\fref{dP1_paths}). We will call $\{ P_i \}$ the set of {\bf reference paths}. We choose these
paths such that they start on face $1$ and end on the specific face. This is possible because
the Beilinson quiver is connected. Then, $\Psi$ maps each of these paths to a Weil divisor (see
\fref{dP1_divs} for the image). These divisors determine a collection of line bundles.

\begin{figure}[ht]
  \epsfxsize = 16cm
  \centerline{\epsfbox{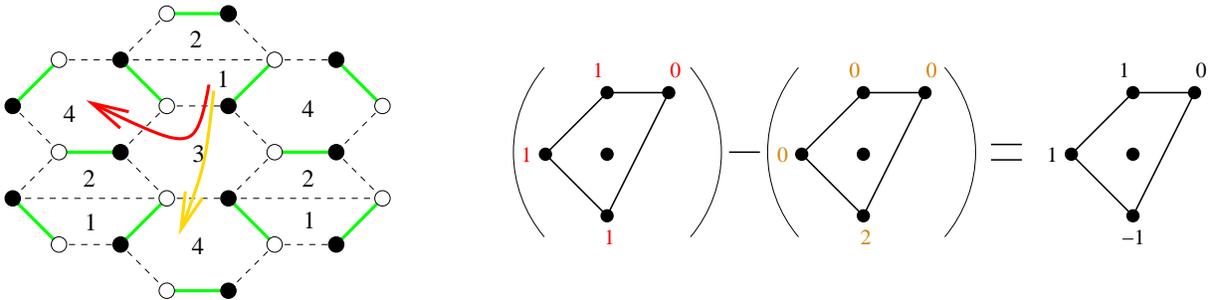}}
  \caption{Face $4$ can be assigned with either the red or the yellow allowed path. The resulting Weil divisors
are shown on the right--hand side. We see that they differ by a linear function, i.e.~they are
equivalent. }
  \label{ambig}
\end{figure}

There is a general freedom in the choice of these paths. The terminal faces can also be chosen
from different fundamental cells. We demonstrate this ambiguity in \fref{ambig}. Let us pick two
different paths that end on the same faces but in different fundamental cells. Recall that
$\Psi$ maps closed loops to linear functions; hence the difference of the resulting divisors is
linear, which means that they are in fact equivalent. Note that $\Psi$ gives the same set of
integers for operators (paths) related by F--term equations.

After determining the divisors that correspond to the $P_i$ paths, we are ready to write down an
exceptional collection. We introduce the notation
\be
   {\mathcal O}(\sum_r a_r D_r) \equiv (a_1, a_2, \dots ,a_n)
\ee
We assign the line bundle of the divisor $\Psi(P_i)$ to the $i^{\textrm{th}}$  face. The integer
numbers sitting at the external nodes are the $a_i$ coefficients. For the first face we assign
$(0,0,\dots,0)$. In our ${\bf dP}_1$ example from \fref{dP1_divs} we obtain the following
collection:
\be
  (0,0,0,0),\ (1,0,0,0),\ (0,0,1,0), \ (1,0,1,1)
  \label{excdp1}
\ee
which is exactly the collection discussed in section \ref{sec:construction}.

Another example for the $Y^{3,2}$ theory is presented in the Appendix.

Before moving on, we would like to point out that the $\Psi$--map efficiently computes the
divisors that correspond to dibaryons. In order to obtain the divisor for the bifundamental $X$,
we simply compute $\Psi(X)$. For ${\bf dP}_1$ we get the following list

{\footnotesize
\begin{center}
\begin{tabular}{|c|c|}
\hline
\ \ \ {\bf field} \ \ \ & \ \ \ {\bf divisor} \ \ \ \\
\hline \hline
$X_{12}$   & $ (1,0,0,0)$ \\
$X_{23}^{(1)},X_{23}^{(2)}$   & $ (0,1,0,0) \cong (0,0,0,1)$ \\
$X_{41}^{(1)},X_{41}^{(2)}$   & $ (0,1,0,0) \cong (0,0,0,1)$  \\
$X_{42}$   & $(0,0,1,0)$ \\
$X_{13}$   & $ (0,0,1,0)$ \\
$X_{34}^{(1)},X_{34}^{(2)},X_{34}^{(3)}$   & $ (0,0,1,0)\cong (1,1,0,0) \cong (1,0,0,1)$ \\
\hline
\end{tabular}
\end{center}
}

\noindent in precise agreement with section 5.1 of \cite{Herzog:2003dj}.  The linear equivalence
relations $\cong$ are easily established.  Let us show that $(0,0,1,0) \cong (1,0,0,1)$. The
difference divisor $(1,0,0,1) - (0,0,1,0) = (1,0,-1,1)$, shown on the right hand side of Figure
\ref{ambig},
 has a $\Psi$ map of the form
$\Psi = y-x$.  In other words $(1,0,-1,1)$ is a principal divisor and the linear equivalence
follows.

In this section we defined the linear $\Psi$--map that computes the divisors corresponding to
the bifundamental fields. This map can be used explicitly to write down a collection of line
bundles for the singularity. Unfortunately, we are lacking a general proof that the generated
collections are always exceptional. Strong exceptionality may be checked on a case--by--case
basis.

\newpage
\subsection{Reconstructing the quiver}
\label{subsec_smatrix}

In section \ref{subsec_psi} we introduced the general method, the $\Psi$--map, that computes a
collection of line bundles that is presumably strongly exceptional. Given such a collection, we
can use rigorous methods to construct the quiver of the gauge theory. In this section we prove
that the quiver obtained this way matches with the dual graph of the tiling which was our
starting point.\footnote{We will prove this for the non--periodic McKay quiver.}

Let us denote the exceptional collection by $\{ E_i \}$.
We define the matrix
\be
  S_{ij} = \mbox{dim Hom}(E_i, E_j).
\ee
The matrix elements in $S$ tell the number of ways of getting from node $i$ to node $j$ in the
Beilinson quiver, taking the relations into account. The inverse of this matrix gives the quiver
directly up to bidirectional arrows. The nonzero elements of $S^{-1}_{ij}$ ($i<j$) are the
number of arrows from $j$ to $i$ minus the number of arrows from $i$ to $j$ in the quiver.

\begin{figure}[ht]
  \epsfxsize = 12cm
  \centerline{\epsfbox{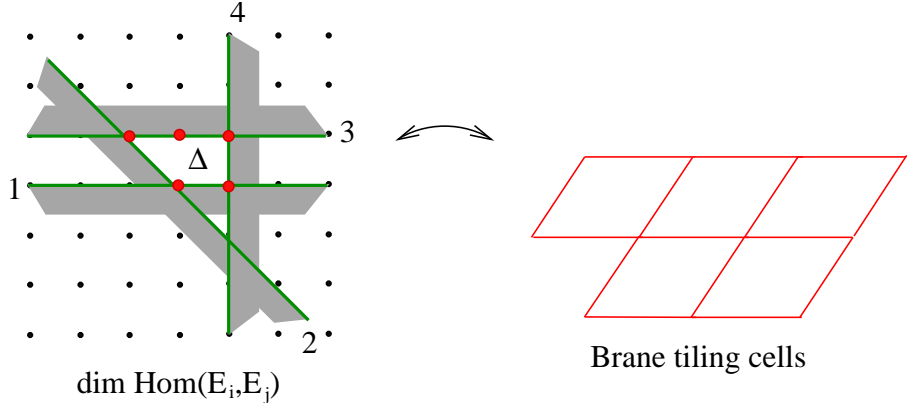}}
  \caption{Determining the $S_{2,4}$ matrix element.
  In this case $E_4 \otimes E_2^* = (1,0,1,1)-(1,0,0,0)=(0,0,1,1)$.
  The figure shows the lattice of the $\Delta_{2,4}$ polygon and its  bounding inequalities.
  The red lattice points inside $\Delta_{2,4}$ can be identified with adjacent fundamental cells in the brane tiling. }
  \label{latticepoints}
\end{figure}

Since we are dealing with line bundles on toric manifolds, the computation of $\mbox{dim
Hom}(E_i, E_j)$ gets vastly simplified \cite{fulton}. This dimension is equal to the number of
global sections of the bundle $E_j \otimes E_i^*$, which we denote by ${\mathcal O}(\sum_r a_r
D_r)$. Then, the dimension is obtained by counting the lattice points inside the polygon
\be
  \Delta_{ij} = \{ {u}\in \IR^2  :  {u}\cdot {v}_r \le a_r \ \mbox{for all } r \}
  \label{deltap}
\ee
where ${v}_r \in \IZ^2$ is the position of the $r$th external node in the toric diagram. See the
left--hand side of \fref{latticepoints} for an example.

In section \ref{subsec_psi} we computed the (\ref{excdp1}) exceptional collection for ${\bf
dP}_1$. Using the above described method, the $S$ matrix and its inverse are determined

\beq
\begin{array}{cc}

  S=\left( \begin{array}{cccc}
  1 & 1 & 3 & 6 \\
  0 & 1 & 2 & 5 \\
  0 & 0 & 1 & 3 \\
  0 & 0 & 0 & 1
  \end{array}
  \right)

& \qquad

  S^{-1}=\left( \begin{array}{cccc}
  1 & -1 & -1 & 2 \\
  0 & 1 & -2 & 1 \\
  0 & 0 & 1 & -3 \\
  0 & 0 & 0 & 1
  \end{array}
  \right)

\end{array}
\eeq
We see that $S^{-1}$ gives precisely the quiver in \fref{quiver_dp1}.

In the following, we will show that this lattice point counting method of determining the number
of paths from node $i$ to node $j$ in the quiver is identical to the same computation on the
brane tiling.  Since the number of paths essentially encodes the quiver via $S$ and $S^{-1}$, we
are proving that the collection of line bundles encodes the quiver of the original brane tiling.

\begin{figure}[ht]
  \epsfxsize = 10cm
  \centerline{\epsfbox{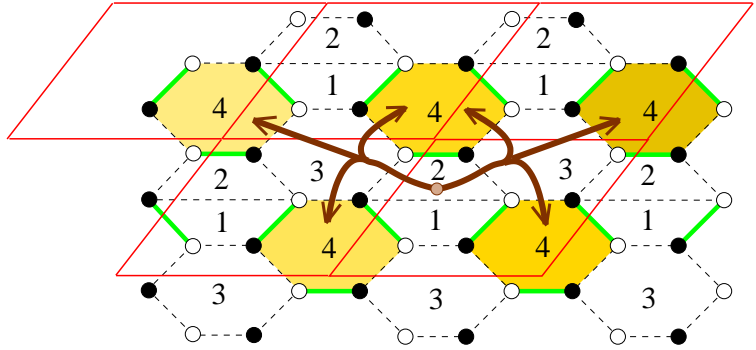}}
  \caption{The figure shows the allowed paths that start on face $2$ and
  end on face $4$. The endpoints of these paths are in different fundamental
  cells which are in one--to--one correspondence with the lattice points inside $\Delta_{2,4}$
  that has been used to compute $\mbox{dim Hom}(E_2, E_4)$.}
  \label{dp1ineq}
\end{figure}

The key observation is that {\it the lattice of $\Delta_{ij}$ can be identified with the lattice
of fundamental cells of the brane tiling}.\footnote{We thank Alastair King for related
discussions.} This is shown in \fref{latticepoints}. In particular, we will assign the lattice
points to the $j^\textrm{th}$ faces in the cells. The simple counting of lattice points also
counts the inequivalent allowed paths from face $i$ to face $j$. There can be many such paths,
but their number is finite, since no loops are allowed. The lattice points in $\Delta_{ij}$ are
in one--to--one correspondence with adjacent fundamental cells that contain the final $j$ faces
where these paths end. In \fref{dp1ineq} these are the five faces marked in yellow. We see that
to one of these faces there are two allowed paths leading. This shouldn't trouble us, since
these are equivalent paths related by the $U_2^1 V^2 = U_2^2 V^1$ F--term equation for the $Y_2$
bifundamental field that separates face $2$ and face $4$. In fact, it turns out that a general
feature of consistent tilings is that homotopic paths of the same length (measured by the
R--charge of the corresponding trace operator\footnote{In fact, any trial R--charge can be used
to measure the length.}) are F--term equivalent. In the following, we will prove this statement.

\bigskip

\begin{figure}[ht]
  \epsfxsize = 4cm
  \centerline{\epsfbox{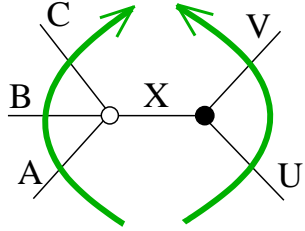}}
  \caption{The F--flatness equation for the $X$ bifundamental field is $C B A = V U$.
  This states the equivalence of the two green paths in the figure.}
  \label{equpf}
\end{figure}

\blemma{In a consistent tiling, paths of the same length are F--term equivalent iff they are
homotopic.} \label{lemma1}
\elemma
\bproof
F--flatness equations are local transformations of the paths (\fref{equpf}); hence they
transform homotopic paths into one another. Applying such a transformation to the path does not
change the R--charge of the corresponding operator. We need to show that two homotopic paths are
equivalent.


\begin{figure}[ht]
  \epsfxsize = 10cm
  \centerline{\epsfbox{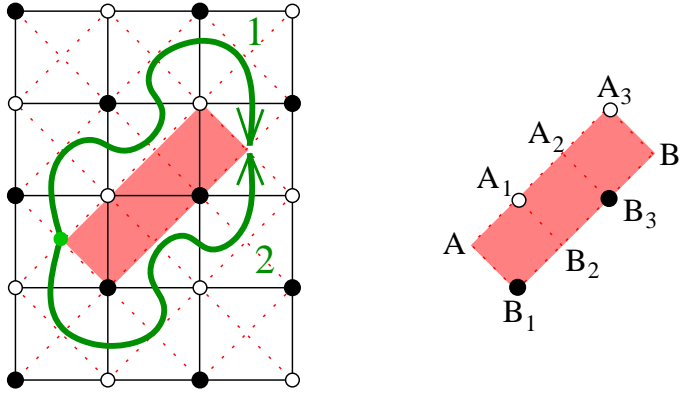}}
  \caption{Homotopic paths are equivalent. The left--hand side of the figure shows two paths
  represented schematically by green lines. The tiling is colored black and the underlying rhombus lattice is shown by
dotted lines. The pink area surrounded by the two paths is also shown separately. }
  \label{equp1}
\end{figure}

As an illustration, \fref{equp1} shows two such paths in a square lattice that can be deformed
into one another by F--terms. The rhombi they surround are also shown separately in the
right--hand side of the figure. This area has two bounding lines: $A A_1 A_2 A_3 B$ and $A B_1
B_2 B_3 B$. On the boundary we find two kinds of rhombus nodes alternating: Every other node is
also a node of the tiling ($A_1, A_3, B_1, B_3$). We call these odd nodes. The remaining even
nodes ($A, A_2, B, B_2$) are only vertices in the rhombus lattice.

We can start deforming path $1$ by using the F--term equation for the tiling edge $A_3 B_3$. We
also see that using the F--term equation for $A_1 B_1$ is not possible because path $1$ does not
contain $A_1 B_3$. At the level of the rhombus lattice the difference of the two nodes $A_3$ and
$A_1$ can be quickly seen: There is no red rhombus lattice edge in the pink area that connects
$A_3$ to another node, whereas $A_1$ has one edge, namely $A_1 B_2$. To summarize, the area
between two paths can be reduced by F--terms where the boundary nodes don't have rhombus edges.

\begin{figure}[ht]
  \epsfxsize = 7cm
  \centerline{\epsfbox{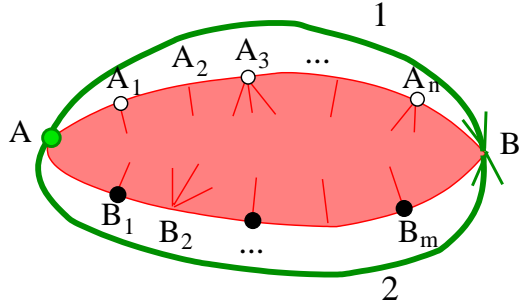}}
  \caption{Two homotopic paths that pass around the pink area. Each boundary node
  ($A_1, \dots, A_n, B_1, \dots, B_m$) has at least
  one rhombus edge which ensures that the area cannot be reduced by F--terms.}
  \label{equp2}
\end{figure}

Let us consider two homotopic paths that start and end on the same two faces. For simplicity, we
assume that the paths are not intersecting. We also assume that the area between the two paths
has been completely reduced, i.e.~there are no more F--terms that we can use to decrease it.
This is equivalent to requiring that the odd nodes along the boundary have at least one rhombus
edge going to the interior of the area. One can check that by construction the even nodes always
have at least one rhombus edge. (In the previous example, such nodes were $A_2$ and $B_2$.) The
reduced area can be schematically drawn as in \fref{equp2}.

\begin{figure}[ht]
  \epsfxsize = 7cm
  \centerline{\epsfbox{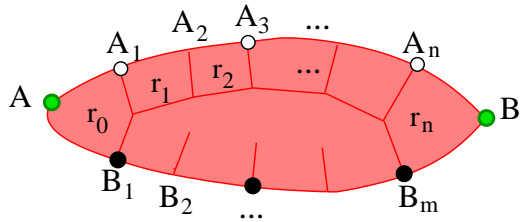}}
  \caption{The straight rhombus path in the area contains rhombi $r_0, \dots , r_n$.
  The existence of this series of rhombi constrains $A B_1$ to be parallel to  $B_m B$.}
  \label{equp3}
\end{figure}

If we suppose that there is precisely one red rhombus edge at each $A_i$ and $B_j$ node and
there are no edges at $A$ and $B$, then we recognize a straight rhombus path built out of the
$r_i$ ($i=0,1,2,\dots,n$) rhombi. These are located at the boundary next to path $1$ (see
\fref{equp3}).

This rhombus path corresponds to a zig--zag path in the tiling. The opposite edges of the rhombi
are parallel; hence $A B_1$ is parallel to $B_m B$. The same argument applies for the rhombi on
the other side of the area; hence $A A_1$ is parallel to $A_n B$. As a consequence, some of the
rhombi in the area must be degenerate (here $r_0$ and $r_n$), i.e.~the R--charges of the
corresponding fields are zero or negative and the tiling is inconsistent. Here we used that
there is one rhombus edge for each node.

Extra rhombus edges joining to $A_i$, $B_j$ or to the endpoints $A$ or $B$ can't be used to
restore the consistency of the tiling since they make the rhombi even more degenerate. This can
also be seen by looking at the sum of internal angles of the $A, A_1, \dots, A_n, B, B_m, \dots,
B_1, A$ pink polygon. This polygon has $n+m+2$ vertices, hence the sum of angles should be
$(n+m)\pi$. Every rhombus next to the boundary contributes $\pi$ to the sum, except for the
rhombi at $A$ and $B$ whose contribution can be bigger. If there are extra rhombus edges at a
particular node, then we also get contribution from those rhombi that touch this node but they
don't have a common edge with the boundary polygon. Since there are at least $n+m$ rhombi, the
total sum of angles is greater than $(n+m)\pi$; hence the polygon must be degenerate. \eproof

As an immediate corollary, the lemma proves the following observation of \cite{Benvenuti:2005cz}

\bcor{The structure of the chiral ring is naturally encoded in the non--trivial cycles of the
tiling torus. In particular, the dual cone can be ``embedded'' in the infinite tiling
\cite{king_vegh}.} \ecor

\begin{figure}[ht]
  \epsfxsize = 12cm
  \centerline{\epsfbox{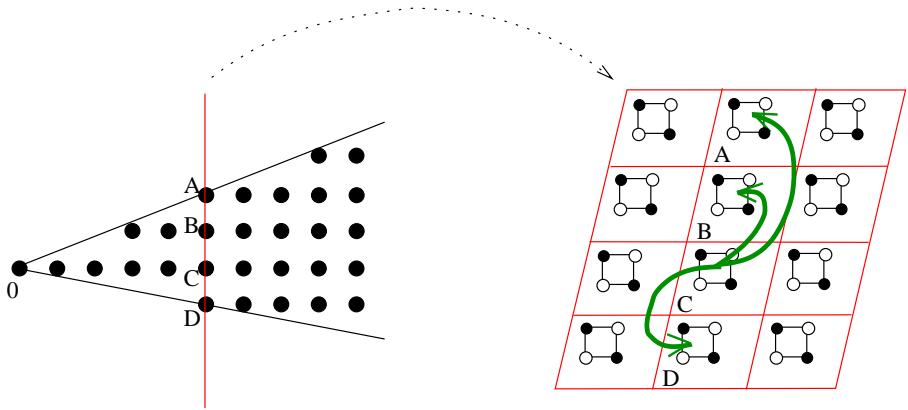}}
  \caption{The embedding of the dual cone in the tiling torus. }
  \label{dualcone}
\end{figure}

The embedding is sketched in \fref{dualcone}.

One can assign gauge invariant mesonic operators to each of the monomials in the dual cone. For
the $A, B, D$ monomials we assigned three green paths that are schematically shown in the
right--hand side of the figure. They start and end on the same square in the tiling. Keeping
these endpoints and the lengths fixed, they can be freely deformed due to Lemma \ref{lemma1}.

Then, the endpoints of the paths in the lattice of fundamental cells can be identified with the
projection of the monomials onto the red tiling plane. To reach the bulk of the cone (here the
monomials $B$ and $C$), the path has to contain loops, e.g.~small loops around a tiling node.
For instance, the tip of the cone and $C$ are projected to the same point; therefore the
corresponding path to $C$ must be a trivial loop. It can be chosen to be the appropriate power
of any term in the superpotential.

\bcor{For the consistency of the tiling a necessary condition is that homotopic paths of the
same length are F--term equivalent.\footnote{ An immediate question arises: Is this condition
sufficient? Can consistency be defined as the equivalence of homotopic paths? We leave this
question for future study.}} \ecor

\begin{figure}[ht]
  \epsfxsize = 4cm
  \centerline{\epsfbox{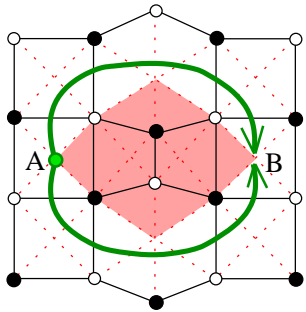}}
  \caption{Inequivalent $A \rightarrow B$ homotopic paths in an inconsistent tiling. }
  \label{inconineq}
\end{figure}

If the tiling is inconsistent, it might be possible to construct two inequivalent paths
surrounding the ``inconsistency''. An example is shown in \fref{inconineq} where the tiling
contains the subgraph of Figure 15 in \cite{Hanany:2005ss}. We recognize the two rhombus paths
and the corresponding tiling zig--zags along the boundary of the pink area. Since no F--terms
can be used, the paths are inequivalent.

 \vskip 0.5cm


After proving the lemma and investigating some of its corollaries, let us turn back to the
original problem. We want to show that the matrix element $S_{ij}$ gives the number of
inequivalent paths from $i$ to $j$. In order to prove this, we need to show that for each ${u}$
lattice point in $\Delta_{ij}$, we have a unique allowed path in the tiling starting on the
$i^{\textrm{th}}$ face and ending on the $j^{\textrm{th}}$ one. These $j^{\textrm{th}}$ faces
are in different fundamental cells that are in one--to--one correspondence with the ${u}$
lattice points.

\begin{figure}[ht]
  \epsfxsize = 11cm
  \centerline{\epsfbox{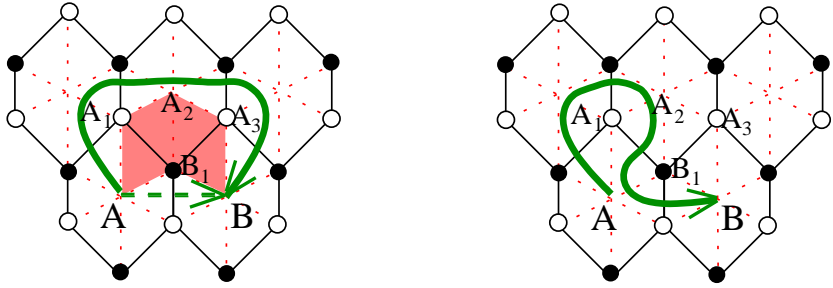}}
  \caption{Homotopic paths with different R--charge are not equivalent.
  After applying the F--term equation for $A_3 B_1$, the long path (solid green line)
  gets transformed to the short path (dashed line) plus a small loop around the $A_1$
  node in the tiling.}
  \label{nonequp}
\end{figure}

The previous lemma ensures that we have a single path for each cell. To see this, we need to
prove that allowed homotopic paths have the same length. Suppose that there exist two homotopic
paths of different lengths. Using F--term equations, we can deform the longer path to the
shorter one as in \fref{nonequp}. Thus, we end up with loops around tiling nodes which are
evidently not allowed, since these loops intersect $PM_0$. Recalling that F--terms transform
allowed paths to allowed paths, we arrive at a contradiction. This means that in a consistent
tiling {\it homotopic allowed paths always have the same R--charge and are equivalent}.

Having proved that from the  $i^{\textrm{th}}$ face of a fixed fundamental cell there exists at
most one inequivalent path to the $j^{\textrm{th}}$ face of any cell, we also need to show that
these cells where the paths can end are in one--to--one correspondence with the ${u}$ lattice
points. In order to do so, we reinterpret the (\ref{deltap}) bounding inequalities of
$\Delta_{ij}$.

In the definition of $\Delta_{ij}$, we have a ${u}\cdot {v}_r \le a_r $ constraint for each
external node of the toric diagram. For a given path, ${u}$ is interpreted as the integer vector
defined on the lattice of fundamental cells giving the distance of the cells wherein the
$i^{\textrm{th}}$ and $j^{\textrm{th}}$ faces reside. In the tiling language, ${v}_r$ is the
monodromy of the height of the $r^{\textrm{th}}$ external perfect matching. Thus, the scalar
product gives the increase in the $r^{\textrm{th}}$ height coordinate. Hence, the $a_r$
variables should be interpreted as height differences. In fact, this is exactly how we computed
them with the $\Psi$--map in section \ref{subsec_psi}.

\begin{figure}[ht]
  \epsfxsize = 14cm
  \centerline{\epsfbox{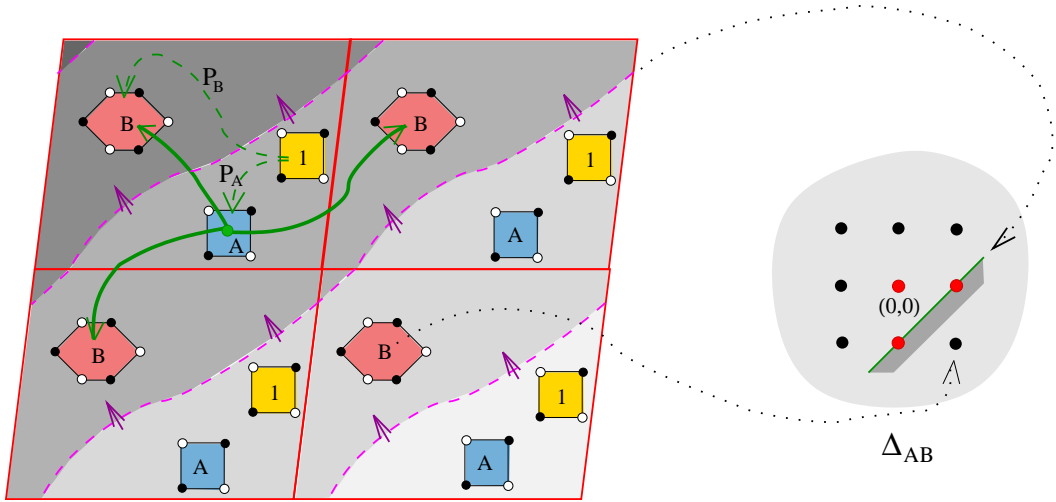}}
  \caption{The figure schematically depicts three allowed green paths from $A$ to $B$.
  The shading indicates one of the height coordinates. The height increases in the direction
  of the small arrows. The allowed paths can only cross the dashed lines in this direction,
  and thus we obtain a bounding inequality for $\Delta_{AB}$. The remaining edges can be determined
  by means of the other heights.}
  \label{final1}
\end{figure}

\fref{final1} illustrates the correspondence schematically. The figure shows three inequivalent
allowed paths that connect face $A$ to different $B$ faces. The shading indicates the
$r^\textrm{th}$ height function. The height changes along the edges in the superposition of the
corresponding matching and $PM_0$. This level set is represented by purple dashed lines.

The right--hand side of \fref{final1} shows the lattice of $\Delta_{AB}$ along with a green
bounding line. The lattice points are in one--to--one correspondence with the red fundamental
cells on the left--hand side. In particular, we assign them to the $B$ faces sitting in the
cells. We set the origin at the middle point which is assigned to the upper left $B$ face in the
tiling.

How does the green constraint come about? From previous discussions in section \ref{subsec_beil}
we know that {\it allowed paths can only go uphill on the height function}. For example, in
\fref{final1} the paths can cross the dashed lines in the direction of the small arrows;
therefore we can't reach the $B$ face in the lower right corner. This face corresponds to the
excluded point on the right--hand side shown by the dotted arrow.

Using the above interpretation of ${u}$, we can immediately write down a necessary (and
sufficient) condition for the allowed paths. In our schematic example, we have ${v}_r = (1,-1)$
which is the average ``gradient vector'' of the height function. Naively, the constraint
translates to the following inequality for the allowed paths
\be
  {u}\cdot {v}_r  \le 0
\ee
This is not quite right, because the paths start from $A$ not $B$. One can take this into
account by adding the difference in their height coordinates to the right--hand side
\be
  {u}\cdot {v}_r  \le d_r
\ee
By using the $P_A$ and $P_B$ reference paths that connect the first node of the Beilinson quiver
to $A$ and $B$, one can see this difference is given by $d_r = \Psi_r(P_B) -
\Psi_r(P_A)$.\footnote{In the example of \fref{final1}, the difference is $d_r = 1-0 = 1$,
i.e.~there is one level line between $A$ and $B$.} Let us denote the $i^\textrm{th}$ line bundle
in the exceptional collection by $(a_1^{i}, a_2^{i}, \dots ,a_n^{i})$. Recalling from section
\ref{subsec_psi} how we have determined the collection, we obtain $d_r = a_r^{B} - a_r^{A}$. Our
final expression is then
\be
  {u}\cdot {v}_r  \le a_r^{B} - a_r^{A}
\ee
which is precisely the inequality in the definition of $\Delta_{AB}$!

We can write down the remaining inequalities for the constraints coming from the other height
functions in exactly the same way. Thus, we obtain the boundaries of $\Delta_{AB}$.

We have seen that the inequalities are equivalent to the fact that allowed paths can't go
downhill on any of the height functions of the external matchings. This completes the
correspondence between the lattice points of $\Delta$ and the allowed paths, and thus proves
that $S_{ij}$ indeed counts the inequivalent paths in the tiling.


\bigskip

Let us summarize the main results of this section. Given a consistent brane tiling, we can
compute a $\mathcal{B}$ Beilinson quiver and an $\{ E_i \}$ collection of line bundles by means
of an internal matching and the $\Psi$--map.\footnote{For a specific Calabi--Yau, there are many
equivalent {\bf Seiberg dual phases} of the quiver theory \cite{Feng:2001xr, Feng:2001bn,
Feng:2002zw, Berenstein:2002fi, Feng:2002kk, Franco:2003ea, Benvenuti:2004wx, Herzog:2004qw,
Hanany:2005ss}. Notice that the exceptional collection of section \ref{subsec_psi} has the
advantage that it gives back the right phase of the theory when computing the $S^{-1}$ quiver
adjacency matrix.} One may check on a case--by--case basis that this collection is exceptional.

In this section we have proved that the ``true'' Beilinson quiver of the gauge theory living in
the worldvolume of the D3--branes is the same as $\mathcal{B}$, the original quiver which is
obtained directly from the tiling. In particular, we proved that the number of inequivalent
paths between two nodes are the same.

As a byproduct, we obtained that homotopic paths with the same R--charge are F--term equivalent.
Thus, we could clarify the relation of the brane tiling to the dual cone by a projection of the
lattice points of the cone onto the tiling plane. This gave an explicit correspondence between
monomials and paths.

\section{Conclusions}
\label{section_conclusions}

Brane tilings can be deceptively simple.  With a few strokes of a pen, all of the data of a
${\mathcal N}=1$ supersymmetric quiver gauge theory -- the matter fields, the gauge groups, the
superpotential -- are captured.  Given these simple pictures, theorems should be easy to prove,
but we have often found otherwise.  In the following paragraphs, we outline our successes but
also the work that remains to be done to prove our dictionary between brane tilings and
exceptional collections.

In section 4, we provided a recipe that will convert any exceptional collection of line bundles
into a periodic quiver and motivated the recipe using Wilson lines and a little mirror symmetry.
In the cases we looked at, this periodic quiver was the graph theoretic dual of a brane tiling.
Thinking of the periodic quiver as a triangulation of a surface, we proved that the Euler
character vanished.   Since the exceptional collection specifies the connectivity of all the
vertices, edges, and faces, a vanishing Euler character is not necessarily enough to ensure the
quiver can be written on a torus.  We hope to return to this issue in the future.

In section 5, we provided a recipe that will convert any brane tiling into a collection of line
bundles.  Two key observations underlie this recipe.  The first is that internal perfect
matchings of the tilling are in one--to--one correspondence with Beilinson quivers and hence
with exceptional collections.  The second is that external perfect matchings are in one--to--one
correspondence with the generating Weil divisors $D_r$ and can be used to convert paths in the
brane tiling into sums of divisors $\sum a_r D_r$ via the $\Psi$--map.

We left the word exceptional out of the first sentence of the preceding paragraph on purpose.  On a case by case basis, we can verify the collections are exceptional, using for example the techniques described in \cite{Herzog:2005sy}.  However, proving that the collection is exceptional in general is difficult.
There is a paper by Altmann and Hille \cite{AltmannHille} who
prove strong exceptionality for quivers without relations (no superpotential) using Kodaira vanishing.
The Kodaira vanishing theorem and certain generalizations are a powerful way
of proving strong exceptionality.  Given a line bundle $\calo(D)$ corresponding to an ample
divisor $D$, then
\be
\dim H^q(X, \calo(D \otimes K))=0 \; , \; \; \mbox{for any } q>0 \ .
\ee
Unfortunately, for us, even in relatively simple exceptional collections, one finds a $D$ which is not ample even though these higher cohomology groups vanish. To see the vanishing, one must rely on techniques specific to the complex surface $V$ in question.

We hope the future brings new progress on both these fronts.

\vskip 0.5cm

{\bf Acknowledgements:} We gratefully acknowledge the invaluable discussions we have had with
Aaron Bergman, Agostino Butti, Charles Doran, Sebastian Franco, Jonathan Heckman, Amer Iqbal,
Robert Karp, Alastair King, David Morrison, Cumrun Vafa and Alberto Zaffaroni. We thank the KITP
for hospitality during part of this work. AH would like to thank the string theory groups in
Milan, Jerusalem, Valdivia and Buenos Aires for their kind hospitality while this work was being
completed.

\section*{Appendix}
\label{appendix}

\begin{figure}[ht]
  \epsfxsize = 7cm
  \centerline{\epsfbox{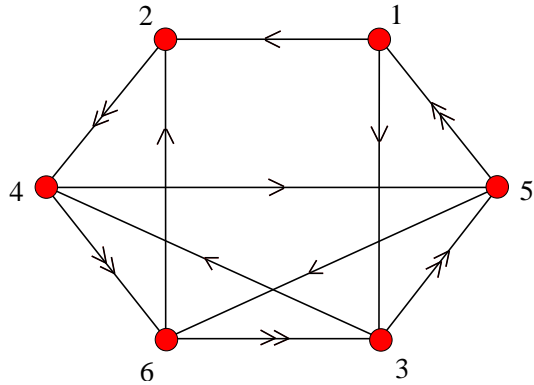}}
  \caption{$Y^{3,2}$ quiver. }
  \label{y32quiver}
\end{figure}

To demonstrate the computation of exceptional collections with the $\Psi$--map of section
\ref{subsec_psi}, we give another example. This is the $Y^{3,2}$ theory, whose quiver is shown
in \fref{y32quiver}.

\begin{figure}[ht]
  \epsfxsize = 15.5cm
  \centerline{\epsfbox{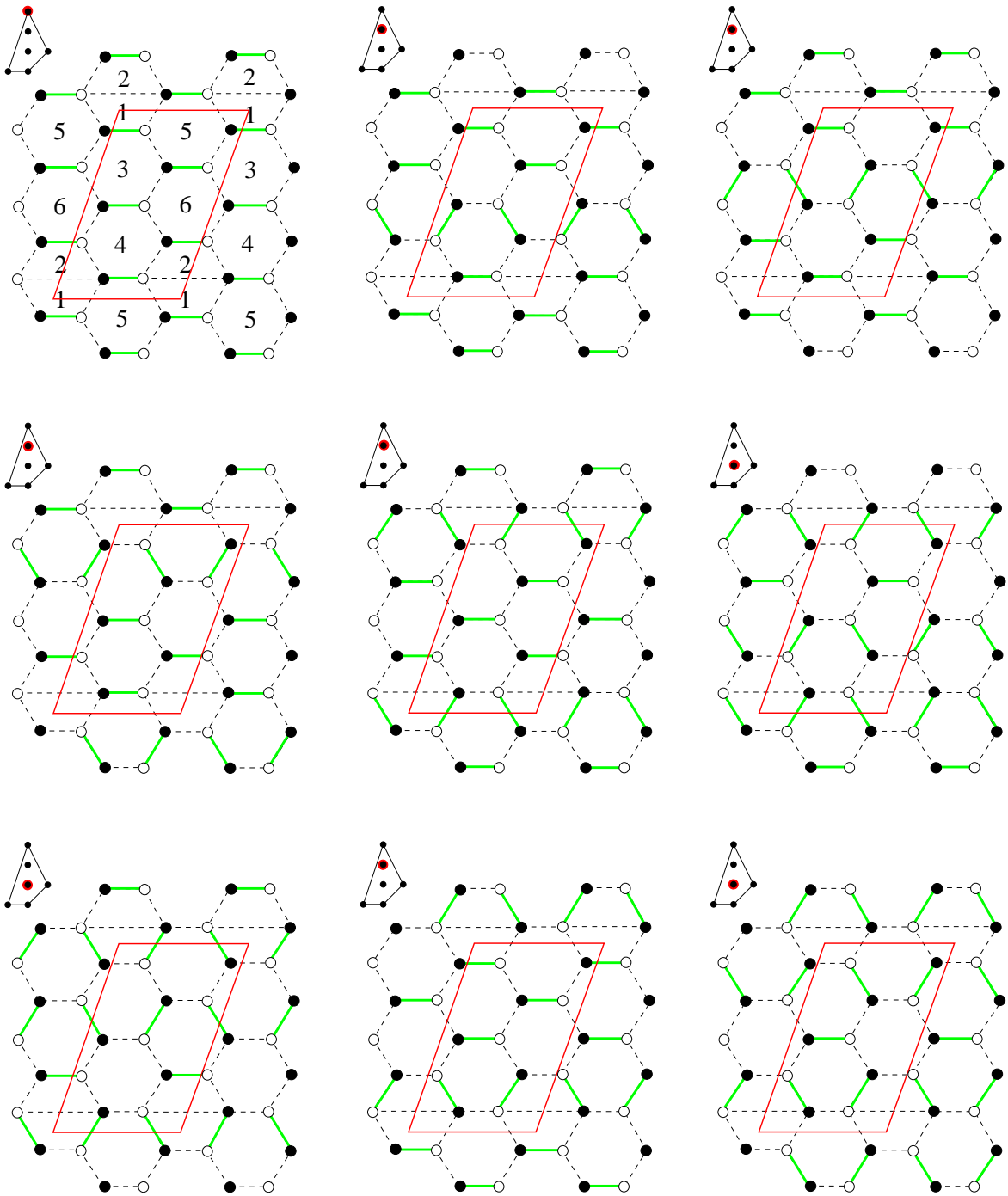}}
  \caption{$Y^{3,2}$ perfect matchings ($1^{st} \dots 9^{th}$). }
  \label{y32pm1}
\end{figure}

\begin{figure}[ht]
  \epsfxsize = 15.5cm
  \centerline{\epsfbox{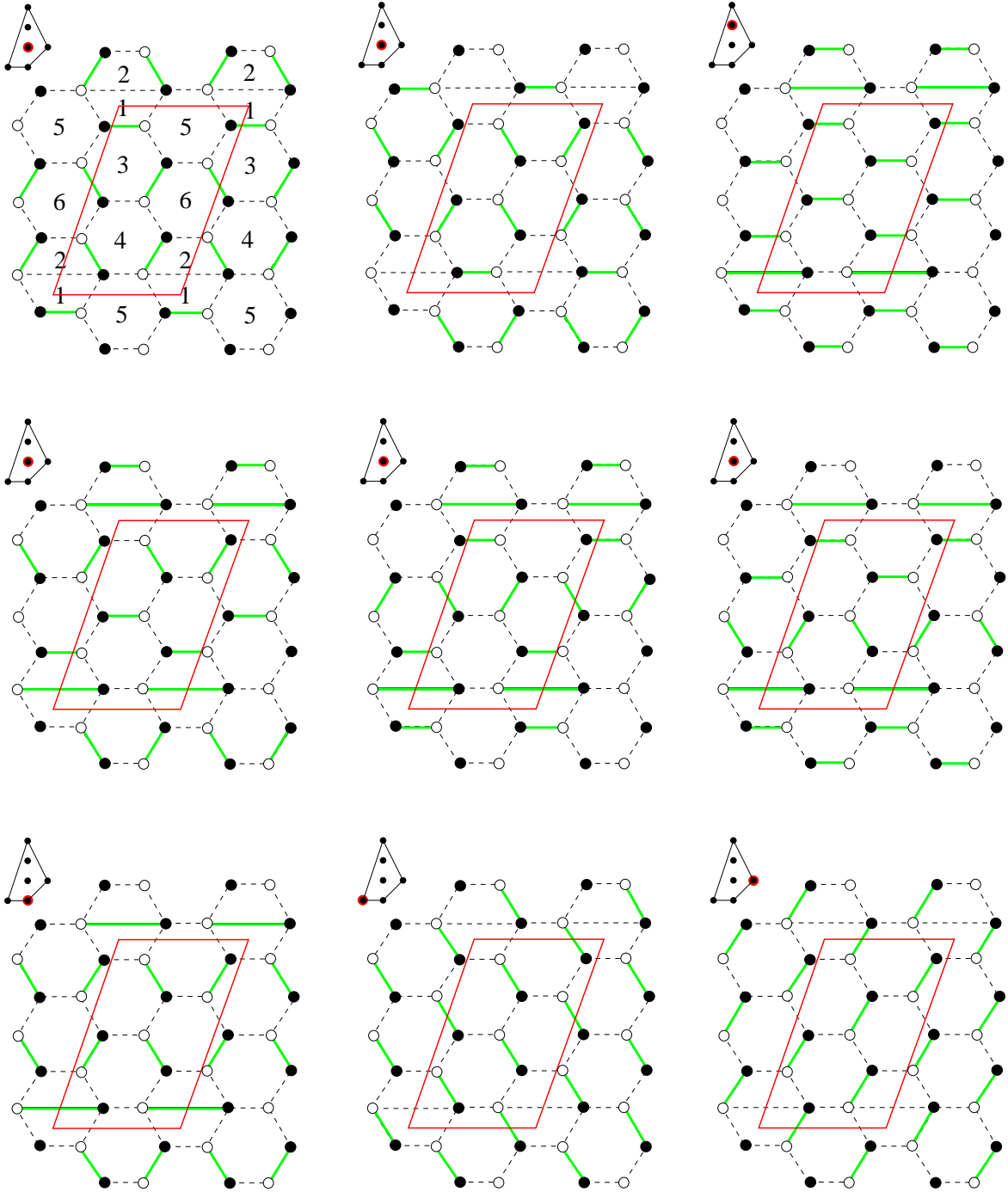}}
  \caption{$Y^{3,2}$ perfect matchings ($10^{th} \dots 18^{th}$). }
  \label{y32pm2}
\end{figure}

\begin{figure}[ht]
  \epsfxsize = 7cm
  \centerline{\epsfbox{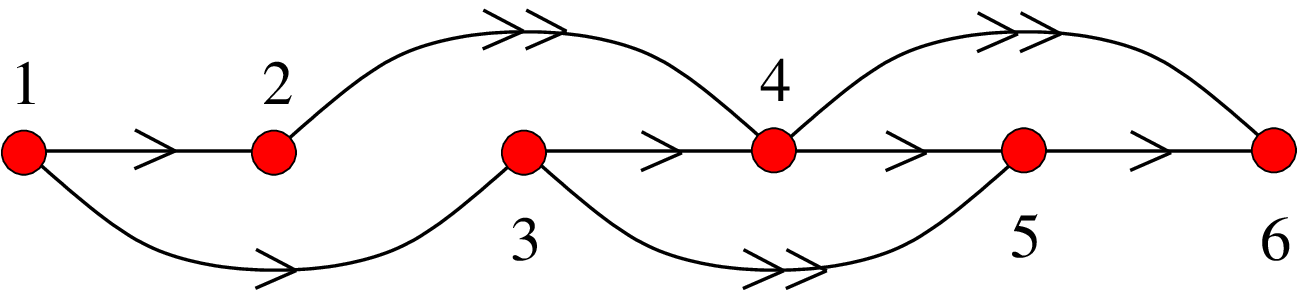}}
  \caption{$Y^{3,2}$ Beilinson quiver. Bifundamentals in internal matching $7$ are omitted.}
  \label{y32beil}
\end{figure}

\begin{figure}[ht]
  \epsfxsize = 5cm
  \centerline{\epsfbox{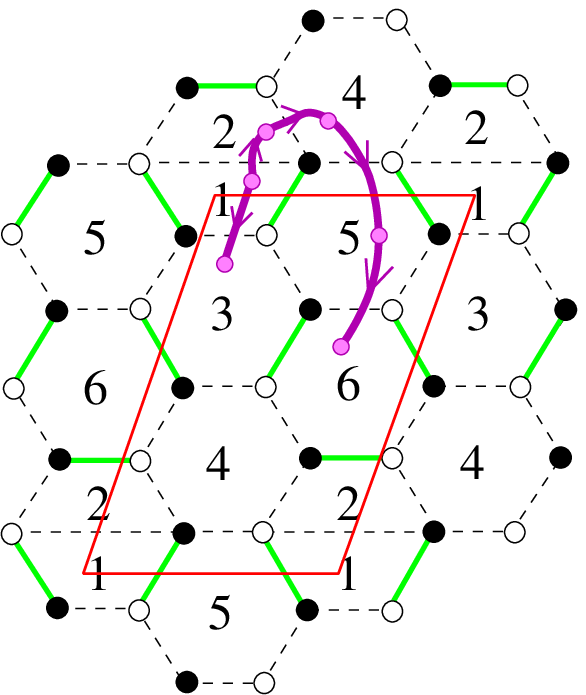}}
  \caption{$Y^{3,2}$ tiling. The purple lines indicate the chosen paths that are used to compute
  the exceptional collections. The paths start on face $1$ and connect it to the other faces.}
  \label{y32paths}
\end{figure}

The brane tiling of this geometry and the 18 perfect matchings are given in \fref{y32pm1} and
\fref{y32pm2}. In the upper left corner of the figures the toric diagram is shown with a red dot
giving the position of the matching. For reference matching we pick the $7^{\textrm{th}}$
matching of \fref{y32pm1}. Deleting the corresponding arrows in the quiver gives the Beilinson
quiver (\fref{y32beil}).  We need to fix allowed reference paths in the tiling that connect the
first node of the Beilinson quiver to all the other nodes. The chosen paths are shown in
\fref{y32paths}.

\newpage

\newpage

\newpage

\begin{figure}[ht]
  \epsfxsize = 13cm
  \centerline{\epsfbox{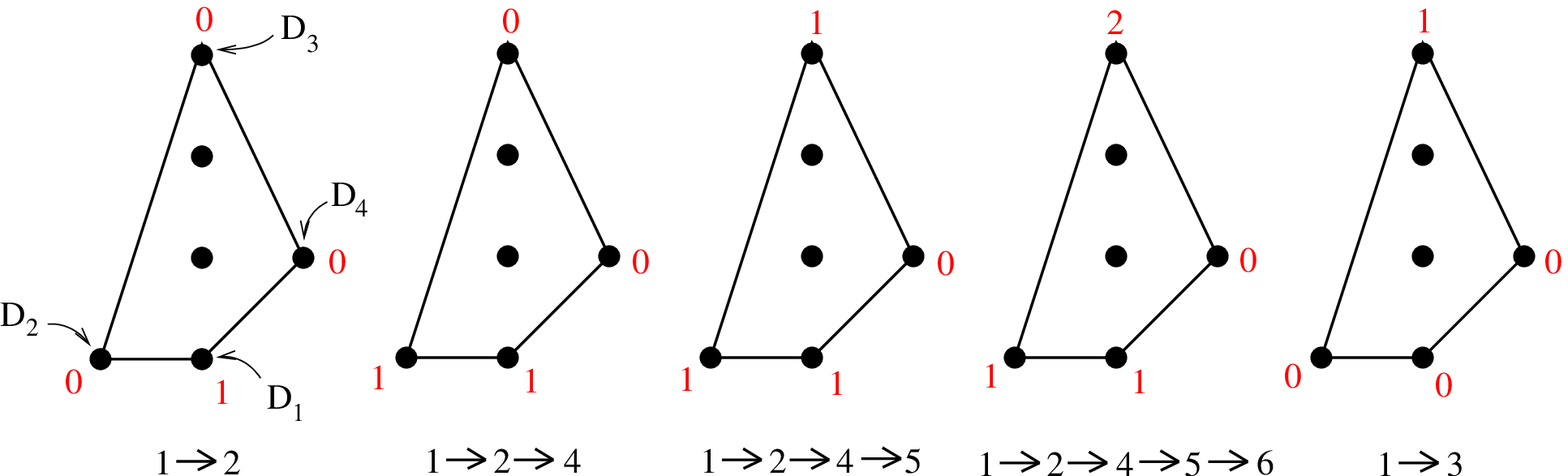}}
  \caption{A set of reference paths for $Y^{3,2}$. }
  \label{y32refpaths}
\end{figure}

From the intersection number of the paths and the external perfect matchings we can immediately
derive the following collection:
\be
  (0,0,0,0),\ (1,0,0,0),\ (0,0,1,0),\ (1,1,0,0),\ (1,1,1,0),\ (1,1,2,0).
\ee


\newpage

\bibliography{paper}
\bibliographystyle{JHEP}

\end{document}